\documentclass [10pt, letterpaper]{article}
\usepackage{fullpage}
\usepackage{amsmath}
\usepackage{amssymb}
\usepackage{graphicx}
\usepackage{mathrsfs}
\usepackage{wrapfig}
\usepackage{boxedminipage}
\usepackage{epsfig}
\usepackage{setspace}
\usepackage{subfigure}
\usepackage{hyperref}
\newcommand{\reef}[1]{(\ref{#1})}

\begin{document}

\begin{flushright}
\phantom{{\tt arXiv:1110.1074}}
\end{flushright}

\bigskip
\bigskip
\bigskip

\begin{center} {\Large \bf Holographic Entanglement Entropy }
  
  \bigskip

{\Large\bf  and }

\bigskip

{\Large\bf      Renormalization Group Flow}

\end{center}

\bigskip \bigskip \bigskip \bigskip

\centerline{\bf Tameem Albash, Clifford V. Johnson}

\bigskip
\bigskip
\bigskip

  \centerline{\it Department of Physics and Astronomy }
\centerline{\it University of
Southern California}
\centerline{\it Los Angeles, CA 90089-0484, U.S.A.}

\bigskip

\centerline{\small \tt talbash,  johnson1,  [at] usc.edu}

\bigskip
\bigskip


\begin{abstract} 
\bigskip
\noindent 
Using holography, we study the entanglement entropy of strongly coupled field theories perturbed by operators that trigger an RG flow from a conformal field theory in the ultraviolet (UV) to a new theory in the infrared (IR). The holographic duals of such flows involve a geometry that has the UV and IR regions separated by a transitional structure in the form of a domain wall.  We address the question of how the geometric approach to computing the entanglement entropy organizes the field theory data, exposing key features as the change in degrees of freedom across the flow, how the domain wall acts as a UV region for the IR theory,  and a new area law controlled by the domain wall.  Using a simple but robust model we uncover this organization, and expect much of it to persist in a wide range of holographic RG flow examples. We test our formulae in two known examples of RG flow in 3+1 and 2+1 dimensions that connect non--trivial fixed points.

\end{abstract}
\newpage \baselineskip=18pt \setcounter{footnote}{0}


%
\section{Introduction}
%
A useful probe of the properties of various field theories that has
received increased interest in recent times is the entanglement
entropy, with applications being pursued in diverse areas such as
condensed matter physics, quantum information, and quantum
gravity. One of the main motivators, in the context of strongly
coupled field theories  (perhaps modeling novel
new phases of matter), is that the entanglement entropy may well act
as a diagnostic of important phenomena such as phase transitions, in
cases where traditional order parameters may not be available.

Within a system of interest, consider a region or subsystem and  call it
$\mathcal{A}$, with the remaining part of the system denoted by
$\mathcal{B}$. A definition of the entanglement entropy of
$\mathcal{A}$ with $\mathcal{B}$ is given by:
\begin{equation}
S_{\mathcal{A}} = - \mathrm{Tr}_{\mathcal{A}} \left( \rho_\mathcal{A} \ln \rho_\mathcal{A} \right)\ ,
\end{equation}
where $\rho_\mathcal{A}$ is the reduced density matrix of $\mathcal{A}$ given by tracing over the degrees of freedom of $\mathcal{B}$,
$\rho_\mathcal{A} = \mathrm{Tr}_{\mathcal{B}}( \rho) $,
%
where $\rho$ is the density matrix of the system.  When the system is in a pure state, \emph{i.e.,}
$\rho = \left| \Psi \rangle \langle \Psi \right|$, 
%
the entanglement entropy is a measure of the entanglement between the
degrees of freedom in $\mathcal{A}$ with those in~$\mathcal{B}$. 

\begin{figure}[h]
\begin{center}
\subfigure[The strip.]{\includegraphics[width=3.0in]{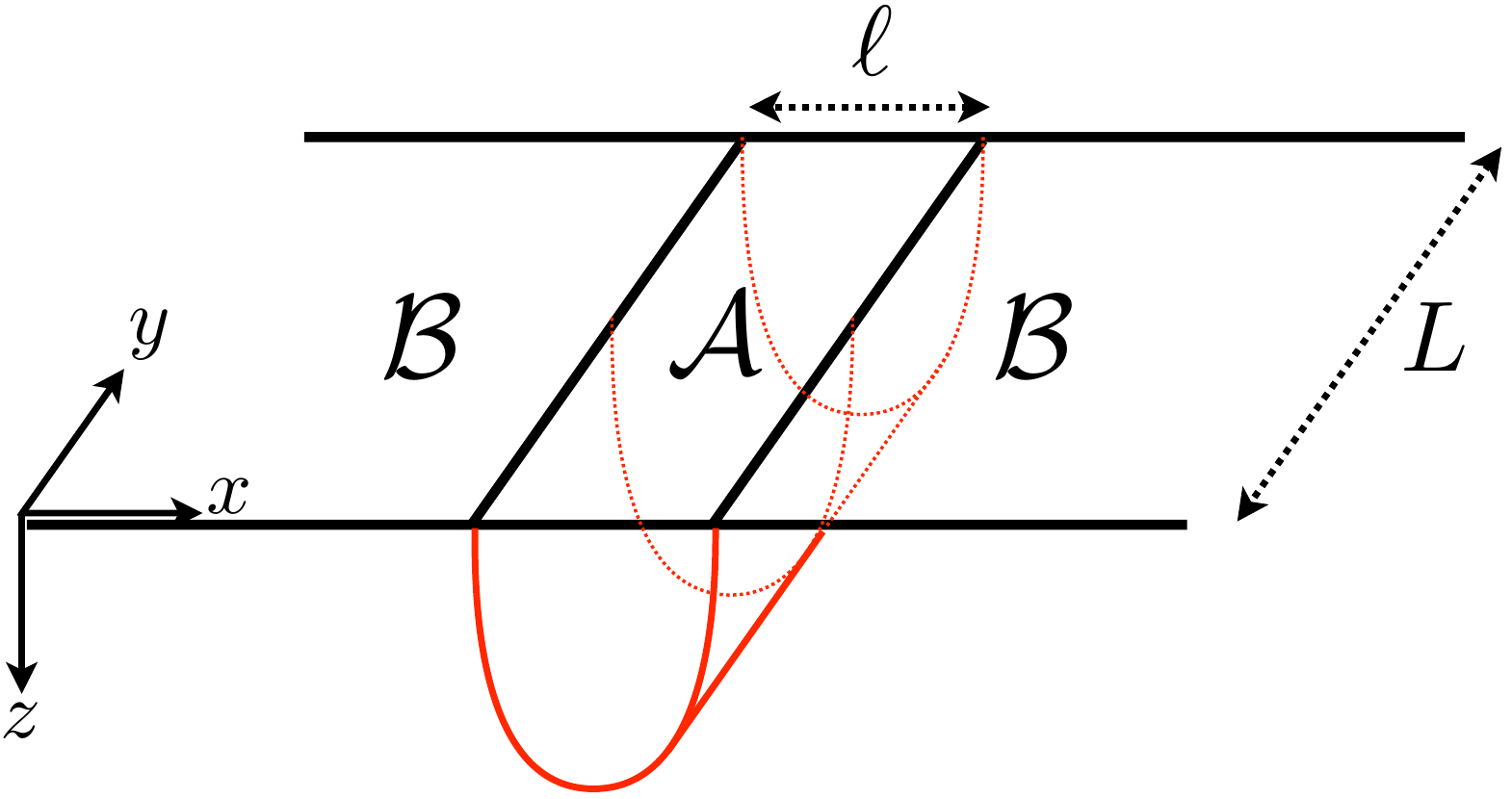}} \hspace{0.5cm}
\subfigure[The disc.]{\includegraphics[width=3.0in]{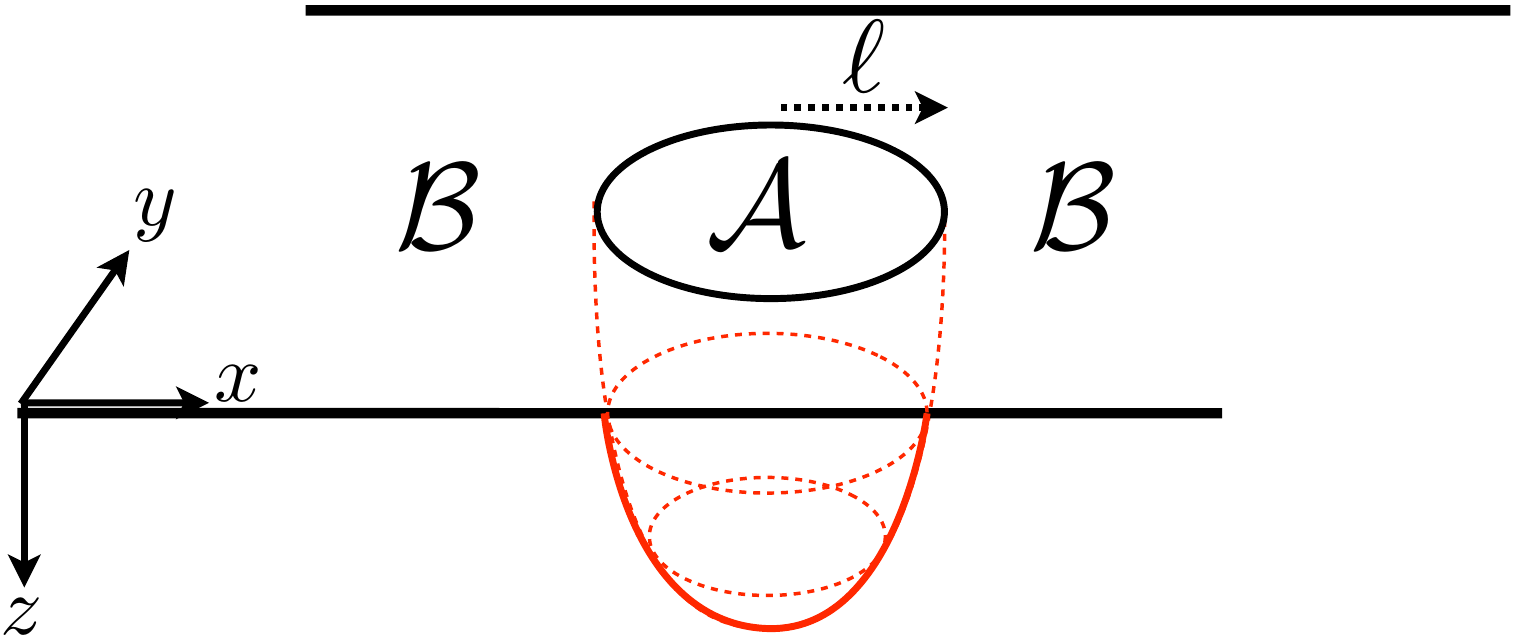}}
   \caption{\small Diagrams of the two shapes we will consider for region $\mathcal{A}$. This is the case of AdS$_4$, and here,  $z$ denotes the radial direction in AdS$_4$. In one dimension higher we will generalize these shapes to a box and a round ball, and in one dimension fewer, we will consider an interval.}  \label{fig:setup_shapes}
   \end{center}
\end{figure}

It is of interest to find ways of computing the entanglement entropy
in various strongly coupled systems, in diverse dimensions, and under
a variety of perturbations, such as the switching on of external
fields, or deformations by relevant operators. A powerful tool for
studying such strongly coupled situations is gauge/gravity duality,
which emerged from studies in string theory and M-theory. The best
understood examples are the conjectured AdS/CFT correspondence and its
numerous deformations
\cite{Maldacena:1997re,Gubser:1998bc,Witten:1998qj,Witten:1998zw} (See \emph{e.g.,} ref.\cite{Aharony:1999t} for
an early, but still very useful, review.) There has
been a great deal of activity for over a decade now, applying these
tools to strongly coupled situations of potential interest in
condensed matter and nuclear physics, for example. Fortunately, there
has been an elegant proposal\cite{Ryu:2006bv,Ryu:2006ef} for how to compute the entanglement
entropy in systems with an Einstein gravity dual (or, more generally, a string or
M--theory dual in the large $N$ limit and large t' Hooft limit), which provides a new way to
calculate the entanglement entropy using geometrical techniques (for a
review see ref.\cite{Nishioka:2009un}). In an asymptotically Anti--de
Sitter (AdS) geometry, consider a slice at constant AdS radial
coordinate $z = a$. Recall that this defines the dual field theory
(with one dimension fewer) as essentially residing on that slice in
the presence of a UV cutoff set by the position of the slice. Sending
the slice to the AdS boundary at infinity removes the cutoff (see
ref. \cite{Aharony:1999t} for a review).  On our $z=a$ slice, consider
a region $\mathcal{A}$. Now find the minimal--area surface
$\gamma_{\mathcal{A}}$ bounded by the perimeter of $\mathcal{A}$ and
that extends into the bulk of the geometry.  (Figure~\ref{fig:setup_shapes} shows examples of the arrangement we will consider in this paper.)
Then the entanglement
entropy of region $\mathcal{A}$ with $\mathcal{B}$ is given by:
\begin{equation}
\label{eq:holographic_entanglement}
S_{\mathcal{A}} = \frac{ \mathrm{Area}(\gamma_{\mathcal{A}})}{4 G_{\rm N}} \ ,
\end{equation}
where $G_{\rm N}$ is Newton's constant in the dual gravity
theory. 

This prescription for the entropy coincides nicely with
various low dimensional computations of the entanglement entropy, and
has a natural generalization to higher dimensional theories.  Note
that there is no formal derivation of the prescription.  Steps have
been made, such as in refs.~\cite{Fursaev:2006ih,Headrick:2010zt}, but
they are not complete.  However, there is a lot of evidence for the
proposal. See \emph{e.g.,}
refs.\cite{Solodukhin:2006xv,Hirata:2006jx,Nishioka:2006gr,Headrick:2007km,Solodukhin:2008dh,Casini:2008as}. A
review of several of the issues can be found in
ref.\cite{Nishioka:2009un}.  Further progress has been made recently
in ref.~\cite{Casini:2011kv}.

In this paper we shall assume that this holographic prescription does
give the correct result for the entanglement entropy in systems with
gravity duals, and proceed to examine the interesting question of how
the entanglement entropy behaves when a system is perturbed by an
operator that triggers a Renormalization Group (RG) flow. For
simplicity, we will work with  flows that connects two conformal field
theories, and we will consider (for concreteness) a four dimensional
example and one in three dimensions. Such examples are  extremely natural to study using holographic duality
since (at large $N$) it is possible to find geometries that represent the full flow
from the maximally supersymmetric  theory to   theories with fewer
super symmetries. (This was first proposed in
refs.\cite{Girardello:1998pd,Distler:1998gb}, and several examples
have since been found.) Flow between field theory fixed points correspond to flows between
fixed points of the supergravity scalar potential. The examples we will study begin with the four
dimensional case of the
flow\cite{Khavaev:1998fb,Freedman:1999gp,Karch:1999pv} to the
Leigh--Strassler point\cite{Leigh:1995ep,Argyres:1999xu}, which
results from giving a mass to one of the ${\cal N}=1$ chiral
multiplets that make up the ${\cal N}=4$ Yang Mills gauge multiplet.
We then  continue with the three dimensional generalization of it
discussed in ref.~\cite{Corrado:2001nv}. The gravity dual of the four
dimensional flow connects AdS$_5\times S^5$ at the $r=+\infty$ extreme
of a radial coordinate $r$ to AdS$_5\times {\cal M}_5$ at $r=-\infty$,
where the space ${\cal M}_5$ results from squashing the $S^5$ along
the flow. There are two of the 42 supergravity scalars switched on at
the latter endpoint, and correspondingly the characteristic radius of
the AdS$_5$ in the IR is larger than that of the UV theory:  The gravity dual for the three dimensional
flow has related structures, this time connecting an AdS$_4\times S^7$
UV geometry to an AdS$_4\times {\cal M}_7$ in the IR, where ${\cal
  M}_7$ results from squashing the $S^7$ along the flow.

Before studying the specific examples, however, we step back and try
to anticipate some of the key physics that we should expect from the
entanglement entropy in this type of situation, more generally. Generically, holographic
RG flow involves a flow from one dual geometry in the UV to another in
the IR, separated by an interpolating region that can be thought of as
a domain wall separating the two regions. The key to understanding the
behaviour of the content of the holographic entanglement entropy
formula is to then understand how the computation incorporates the
structure of the domain wall, and how the field theory quantities it
extracts are encoded. To anticipate how to mine this information, we
do an analytic computation of the proposed entanglement
entropy~\reef{eq:holographic_entanglement} in an idealized geometry
given by a sharp domain wall separating two AdS regions with different
values for the cosmological constant.  Working in various dimensions
(AdS$_5$, AdS$_4$, and AdS$_3$, pertaining to flows in four, three,
and two dimensional field theories), we find a fascinating and
satisfying structure, seeing how the entanglement entropy tracks the
change in degrees of freedom under the flow, and several other features. We
expect that these features will be present in a wide range of
examples, and we confirm our results in the examples mentioned above.

The outline of this paper is as follows. In section~2 we carry out the
study of the entanglement entropy in the presence of the idealized (\emph{i.e.,}  sharp domain wall)
holographic RG flow model, and discover how the physics is organized
in the results. Then, ready to study examples, we review the four
dimensional Leigh--Strassler RG flow of interest, and its dual AdS$_5$
flow geometry in section~3. We explicitly solve (numerically) the
non--linear equations that define the geometry and scalars in the
interpolating dual supergravity flow.  We then compute the
entanglement entropy and extract the physics, comparing to our
predictions from section~2.  Section~4 presents the analogous studies
for the three dimensional field theory, with the AdS$_4$ dual flow
geometry.  We end with  a
 discussion in section 5.

\section{Entanglement Entropy and a Sharp Domain Wall Model}

As mentioned in the introduction, the generic holographic RG flow
involves a flow from one dual geometry in the UV to another in the IR,
separated by an interpotating domain wall.
In all examples, understanding the behaviour of holographic
entanglement entropy, as proposed in
equation~\reef{eq:holographic_entanglement}, requires us to understand
how the area formula incorporates the structure of the domain wall in terms of field theory quantities.  So we start by
doing an analytic computation in an idealized geometry given by a
sharp domain wall in AdS. In general, the location of the wall, and
its thickness, are determined by field
theory parameters corresponding to the details of the relevant
operator - for example, in the case of the Leigh--Strassler flow and
its generalization we later study, the detail in question is the bare
value of the mass given to  the chiral  multiplet. A sharp
domain wall is of course not a supergravity solution, and falls
somewhat outside the usual supergravity duality to any (large~$N$)
theory, but nevertheless is a clean place to start to capture how the
physics is organized. We expect it to capture a great deal of the
key physics of holographic RG flow, as regards how the entanglement
entropy formula works.

We use the following background metric:
\begin{equation} 
ds^2 = e^{2 A(r)} \left( - dt^2 + d \vec{x}^2 \right) + dr^2 \ ,
\end{equation}
with
\begin{equation}
A(r) = \left\{ \begin{array} {lr}
r / R_{\rm UV}  \ , & r > r_{DW}\\
r / R_{\rm IR} \ , & r < r_{DW}
\end{array} \right. \ .
\end{equation}
Here $-\infty<r<+\infty$, and $\vec{x}$ is either four, three, or two
coordinates (the spatial coordinates of the dual field theory),
depending upon whether we are in AdS$_5$, AdS$_4$, or AdS$_3$, the
cases we will consider. Also, $R_{\rm IR}>R_{\rm UV}$. 
The length scale of AdS
on either aide of the wall is set by $R_{\rm UV}$ in the UV at $r>r_{DW}$
and $R_{\rm IR}$ in the IR at $r<R_{DW}$.

\subsection{The Ball and AdS$_5$.}

We begin by studying a region $\mathcal{A}$ in the three spatial
dimensions which is a round ball of radius $\ell$. Using a radial
coordinate $\rho$ in the spatial dimensions, the area of the surface,
$\gamma$, that extends into the bulk is given by:
\begin{equation} \label{eqt:area}
\mathrm{Area} = 4 \pi \int_0^\ell d \rho \ \rho^2 e^{3 A(r)} \left( 1 + e^{-2 A(r)} r'(\rho)^2 \right)^{1/2}\ ,
\end{equation}
where the function $r(\rho)$ defines the enbedding.  We can calculate the equations of motion that result  from minimizing this ``action,'' and we find that the solution is given by:
\begin{equation} \label{eqt:r_rho}
r (\rho) = \left \{ \begin{array} {lr}
- \frac{R_{\rm UV}}{2} \ln \left( \frac{\ell^2 + \epsilon^2 - \rho^2}{R_{\rm UV}^2} \right) \ , & \rho \geq \rho_{DW} \\
- \frac{R_{\rm IR}}{2} \ln \left( \frac{ \ell^2 + \epsilon^2 - \rho^2 + R_{\rm IR}^2 e^{-2 \frac{r_{DW}}{R_{\rm IR}}} - R_{\rm UV}^2 e^{-2 \frac{r_{DW}}{R_{\rm UV}}}}{R_{\rm IR}^2} \right) \  , & \rho < \rho_{DW}
\end{array} \right.
\end{equation}
where $r_{DW}$ is the position of the domain wall in the AdS radial
direction, and $\rho_{DW}$ is the spatial radial position where
$r(\rho_{DW}) = r_{DW}$ and is given by:
\begin{equation}
\rho_{DW}^2 = \ell^2 + \epsilon^2 - R_{\rm UV}^2 e^{-2 \frac{r_{DW}}{ R_{\rm UV}}} \ .
\end{equation}
Note that, rather than integrating out to $r=+\infty$, we integrate
out to large positive radius $r_{\rm UV}$, defining our UV cutoff, with
small $\epsilon$ defined by:
\begin{equation}
r_{\rm UV} = - R_{\rm UV} \ln \left( \frac{\epsilon}{R_{\rm UV}} \right)\ .
\end{equation}
Note that with the solution given for $\rho(\rho)$, we are assuming
that $\ell$ is larger than a critical radius $\ell_{cr}$ such that our
surface extends past the doman wall into the second AdS region.  The
critical radius $\ell_{cr}$ is given by setting $\rho_{DW} =0$ in the
above:
\begin{equation}
\ell_{cr}^2 = R_{\rm UV}^2 e^{-2 \frac{r_{DW}}{R_{\rm UV}}} - \epsilon^2\ .
\end{equation}
Substituting the solution back into equation \reef{eqt:area}, we can
analytically calculate the area of our minimal surface, $\gamma_{\cal
  A}$, and hence the entanglement entropy {\it via}
equation~\reef{eq:holographic_entanglement}. This gives a long
expression that we will not display here. For our purposes it is
enough to first expand the area for small $\epsilon$:
\begin{eqnarray}
\frac{\mathrm{Area}}{4 \pi} &=& \frac{R_{\rm UV}^3}{2}\left[ \frac{\ell^2}{\epsilon^2} + \ln \left( \frac{\epsilon}{\ell} \right)\right] + \frac{R_{\rm UV}^3 }{4}\left[1 - 2  \ln(2)\right]  \nonumber \\
&& - \frac{1}{2} e^{2 \frac{r_{DW}}{R_{\rm UV}}} R_{\rm UV} \ell \sqrt{ \ell^2 - R_{\rm UV}^2 e^{-2 \frac{r_{DW}}{R_{\rm UV}}}} + \frac{1}{2} R_{\rm UV}^3 \tanh^{-1} \left( \frac{\sqrt{ \ell^2 - R_{\rm UV}^2 e^{-2 \frac{r_{DW}}{R_{\rm UV}}}}}{\ell} \right)  \nonumber \\ 
&& + \frac{R_{\rm IR}}{2} e^{2 \frac{r_{DW}}{R_{\rm IR}}} \sqrt{ \ell^2 - R_{\rm UV}^2 e^{-2 \frac{r_{DW}}{R_{\rm UV}}}} \sqrt{  \ell^2  + R_{\rm IR}^2 e^{-2 \frac{r_{DW}}{R_{\rm IR}}} - R_{\rm UV}^2 e^{-2 \frac{r_{DW}}{R_{\rm UV}}}  } \nonumber  \\
&& - \frac{R_{\rm IR}^3}{2} \tanh^{-1} \left( \sqrt{ \frac{\ell^2 - R_{\rm UV}^2 e^{-2 \frac{r_{DW}}{R_{\rm UV}}}}{ \ell^2  + R_{\rm IR}^2 e^{-2 \frac{r_{DW}}{R_{\rm IR}}} - R_{\rm UV}^2 e^{-2 \frac{r_{DW}}{R_{\rm UV}}} } } \right)+O(\epsilon)\ .
\end{eqnarray}
We find  it useful to rewite it in a suggestive way:
\begin{eqnarray} \label{eqt:test1}
\frac{\mathrm{Area}}{4 \pi} &=& \frac{R_{\rm UV}^3}{2}\left[ \frac{\ell^2}{\epsilon^2} + \ln \left( \frac{\epsilon}{\ell} \right)\right] + \frac{R_{\rm UV}^3 }{4}\left[1 - 2  \ln(2)\right]  \nonumber  \\
&& - \frac{1}{2} e^{2 \frac{r_{DW}}{R_{\rm UV}}} R_{\rm UV} \ell \sqrt{ \ell^2 - \tilde{\ell}_{cr}^2} + \frac{1}{2} R_{\rm UV}^3 \tanh^{-1} \left( \frac{\sqrt{ \ell^2- \tilde{\ell}_{cr}^2}}{\ell} \right)  \nonumber \\ 
&& + \frac{R_{\rm IR}}{2} e^{2 \frac{r_{DW}}{R_{\rm IR}}} \sqrt{ \ell^2 - \tilde{\ell}_{cr}^2} \sqrt{  \ell^2 - \tilde{\ell}_{cr}^2 + R_{\rm IR}^2 e^{-2 \frac{r_{DW}}{R_{\rm IR}}}}   \nonumber  \\
&& - \frac{R_{\rm IR}^3}{2} \tanh^{-1} \left( \sqrt{ \frac{\ell^2 - \tilde{\ell}_{cr}^2}{ \ell^2  - \tilde{\ell}_{cr}^2 + R_{\rm IR}^2 e^{-2 \frac{r_{DW}}{R_{\rm IR}}} }}  \right) +O(\epsilon)\ ,
\end{eqnarray}
where
\begin{equation} \label{eqt:tildeellcr}
 \tilde{\ell}_{cr}^2 = R_{\rm UV}^2 e^{-2 \frac{r_{DW}}{R_{\rm UV}}} = \ell_{cr}^2 + O(\epsilon^2)\ .
 \end{equation}
 There are a number of notable features of this expression. First, we
 see the results from pure AdS$_5$ in the first line. There, we see
 the usual UV divergent terms and the $\ell$--independent constant
 that results from the fact that the ball preserves some of the
 conformal invariance of AdS$_5$. Second, the terms that have $R_{\rm IR}$ as
 coefficients (the last two lines) always have $\ell^2$ appearing in
 the combination:
\begin{equation}
\tilde{\ell}^2 = \ell^2 - R_{\rm UV}^2 e^{-2 \frac{r_{DW}}{R_{\rm UV}}} = \ell^2 - \ell_{cr}^2 + O(\epsilon^2)\ .
\end{equation}
We are tempted to interpret this $\tilde\ell$ as the effective ball
radius as seen in the IR, as opposed to the simple $\ell$ seen in the
UV.  Furthermore, the combination:
\begin{equation} \label{eqt:tildeell}
\tilde{\epsilon} = R_{\rm IR} e^{- \frac{r_{DW}}{R_{\rm IR}}} 
\end{equation}
appears in a manner analogous to how the UV cut-off $\epsilon$
appears. (This might not be clear in our $\epsilon$ expansion of the
above equation.  One way to see that it does appear as $\epsilon$
does is to look at equation \reef{eqt:r_rho}).  Now $\tilde{\epsilon}$
is not necessarily small, but we will see that it is useful to think
of it as the cut--off in the IR theory.  With these observations in
mind, we rewrite the last three lines of equation \reef{eqt:test1} as
follows:
\begin{equation}\label{eqt:threelines}
- \frac{R_{\rm UV}^3}{2} \frac{\ell \tilde{\ell}}{\tilde{\ell}_{cr}^2} + \frac{1}{2} R_{\rm UV}^3 \tanh^{-1} \left( \frac{\tilde{\ell}}{\ell} \right) + \frac{R_{\rm IR}^3}{2}  \frac{\tilde{\ell}^2}{\tilde{\epsilon}^2} \sqrt{ 1 - \frac{\tilde{\epsilon}^2}{\tilde{\ell}^2}} - \frac{R_{\rm IR}^3}{2} \tanh^{-1} \left( \frac{1}{\sqrt{1 +  \frac{\tilde{\epsilon}^2}{\tilde{\ell}^2}}} \right)\ .
\end{equation}
Let us focus on the terms proportional to $R_{\rm IR}^3$. If we expand
these terms assuming that $\tilde{\epsilon} / \tilde{\ell} << 1$,
{\it i.e.,} the effective length in the putative IR theory is larger than the IR
cutoff,  which also means that the length is such that the surface
extends very far past the wall into the IR AdS space, we get:
\begin{equation}
\frac{R_{\rm IR}^3}{2}\left[ \frac{\tilde{\ell}^2}{\tilde{\epsilon}^2} +  \ln \left( \frac{\tilde{\epsilon}}{\tilde{\ell}} \right)\right] + \frac{R_{\rm IR}^3}{4} \left[ 1 - 2\ln(2) \right] + O(\tilde{\epsilon}/\tilde{\ell})\ .
\end{equation}
Pleasingly, this is exactly the result we would have obtained  if we
were purely in the IR theory!

So far therefore, we have seen how the entanglement entropy formula
encodes key behaviours of both the UV and the IR theories, in terms of
the appropriate scales, $\epsilon/\ell$ and ${\tilde
  \epsilon}/{\tilde\ell}$. The boundary of AdS$_5$ at $r=+\infty$ is
the UV region and the quantities of the UV theory appear
accordingly. From the point of view of the IR theory, the domain wall
acts (for ${\tilde \epsilon}/{\tilde\ell}$ small) as the effective UV
region, with ${\tilde \epsilon}/{\tilde\ell}$ acting as the effective
regulator.

We are left with understanding the first two terms in
equation~\reef{eqt:threelines}.
These two terms mix the properties of the UV and the IR regions, and
are more subtle.  We associate them with the region around the domain
wall, which connects the UV and IR regions (through and abrupt change
in our idealized example). It is prudent to try to understand the role
of these terms toward the end of the flow, and so we do a large $\ell$
expansion of them, giving:
\begin{equation}
  - \frac{R_{\rm UV}^3}{2} \frac{\ell \tilde{\ell}}{\tilde{\ell}_{cr}^2} + \frac{1}{2} R_{\rm UV}^3 \tanh^{-1} \left( \frac{\tilde{\ell}}{\ell} \right)  = - \frac{R_{\rm UV}^3}{2}\left[ \frac{\ell^2}{\tilde{\ell}_{cr}^2}  +  \ln \left( \frac{\tilde{\ell}_{cr}}{\ell} \right)\right] + \frac{R_{\rm UV}^3}{4}  \left[1   + 2  \ln(2) \right] + O(1 / \ell)\ .
\end{equation}
So we see that these terms give contributions very analogous to our UV
and IR results, where here the reference scale is played by
$\tilde{\ell}_{cr}$.  (Note that the constant term is actually
different than the UV and IR constant terms' form.) At fixed
$\epsilon$ or $\tilde\epsilon$, we may think of this as a new set of
divergences.

Now that we have an understanding of the contributions of the various
pieces to the area, we combine everything together again and consider
the large $\ell$ (and small $\epsilon$) expansion:
\begin{eqnarray}
\frac{\mathrm{Area}}{4 \pi} &=& \frac{R_{\rm UV}^3}{2}\left[ \frac{\ell^2}{ \epsilon^2} + \ln\left( \frac{\epsilon}{\ell}\right)\right] + \frac{\ell^2}{2} \left( \frac{R_{\rm IR}^3}{\tilde{\epsilon}^2} - \frac{R_{\rm UV}^3}{\tilde{\ell}_{cr}^2}  \right) + \frac{R_{\rm UV}^3}{2} \ln \left( \frac{ \ell}{\tilde{\ell}_{cr}} \right) - \frac{R_{\rm IR}^3}{2} \ln \left( \frac{ \ell}{\tilde{\epsilon}} \right)  \nonumber \\
&&  + \frac{R_{\rm UV}^3}{2} + \frac{R_{\rm IR}^3}{4} - \frac{1}{2} R_{\rm IR}^3 \frac{\tilde{\ell}_{cr}^2}{\tilde{\epsilon}^2}   - \frac{1}{2}  R_{\rm IR}^3  \ln(2) + O\left(\frac{1}{\ell} , \ \epsilon \right) \nonumber \\
&=& \frac{R_{\rm UV}^3}{2}\left[ \frac{\ell^2}{ \epsilon^2} + \ln\left( \frac{\epsilon}{\tilde{\ell}_{cr}}\right)\right] + \frac{\ell^2}{2} \left( \frac{R_{\rm IR}^3}{\tilde{\epsilon}^2} - \frac{R_{\rm UV}^3}{\tilde{\ell}_{cr}^2}  \right)  - \frac{R_{\rm IR}^3}{2} \ln \left( \frac{ \ell}{\tilde{\epsilon}} \right)  \nonumber \\
&&  + \frac{R_{\rm UV}^3}{2} + \frac{R_{\rm IR}^3}{4} - \frac{1}{2} R_{\rm IR}^3 \frac{\tilde{\ell}_{cr}^2}{\tilde{\epsilon}^2}   - \frac{1}{2}  R_{\rm IR}^3  \ln(2) + O\left(\frac{1}{\ell} , \ \epsilon \right)  \ . \label{eqt:finalexpand5}
\end{eqnarray}
The first key result here is that we no longer have a $\ln(\ell)$
scaling associated with the UV theory.  The remaining $\ln(\ell)$
dependence has a coefficient that is only associated with the IR
theory and that is independent of the domain wall.  In a non--RG flow scenario, the coefficient of such a term is determined by the central charge of
the theory (see \emph{e.g.,} refs.\cite{Ryu:2006ef,Gubser:1998vd}), but here we see that the coefficient has shifted from its UV
value (associated with the UV central charge) to its IR value (associated with the IR central charge).
The second thing to note is that the area law associated with the UV
cut--off (the first $\ell^2$ term) is joined by a second area law. Its
coefficient is sourced by the details of the domain wall.  For
clarity, we display this term here:
\begin{equation}\label{eqt:new_area_law}
 \frac{\ell^2}{2} \left( \frac{R_{\rm IR}^3}{\tilde{\epsilon}^2} - \frac{R_{\rm UV}^3}{\tilde{\ell}_{cr}^2}  \right) =  \frac{\ell^2}{2} \left( R_{\rm IR} e^{2 \frac{r_{DW}}{R_{\rm IR}}} - R_{\rm UV} e^{2 \frac{r_{DW}}{R_{\rm UV}}}  \right) \ .
\end{equation}
We expect this new area law to be a robust feature of RG flow
geometries, but anticipate that the coefficient's precise form will be
different as we move away from the thin wall limit we are in here.
The above result predicts that the coefficient grows more positive as
$r_{DW}$ is pushed to the UV. In realistic RG flows, while the domain
wall position and sharpness cannot be varied arbitrarily, it is
expected to get thinner toward the UV and so at least in that regime
we should recover positivity.
Finally, the constant terms in the last line of
equation~\reef{eqt:finalexpand5} are a mixture of both the UV, IR, and
domain wall physics.

\subsection{The Disc and AdS$_4$.}
%
We can repeat the same procedure for AdS$_4$, pertaining to RG flows
in $2+1$ dimensional theories.  As our system ${\cal A}$ we consider a
circular disc of radius $\ell$. The solution for the surface embedding
are exactly as in equation \reef{eqt:r_rho}.  We can calculate the minimal 
area and expand for small $\epsilon$ to get:
\begin{equation}
\frac{\mathrm{Area}}{2 \pi} = R_{\rm UV}^2 \frac{\ell}{\epsilon}  - R_{\rm IR}^2 - R_{\rm UV} e^{\frac{r_{DW}}{R_{\rm UV}}} \ell + R_{\rm IR} e^{\frac{r_{DW}}{R_{\rm IR}}} \sqrt{ \ell^2 - R_{\rm UV}^2 e^{-2 \frac{r_{DW}}{R_{\rm UV}}} + R_{\rm IR}^2 e^{2 \frac{r_{DW}}{R_{\rm IR}}}  } + O(\epsilon)\ .
\end{equation}
We see the reappearance of many of the key players that we saw in the
AdS$_5$ case, such as $\tilde{\ell}$, $\tilde{\ell}_{cr}$ and
$\tilde{\epsilon}$, appearing in similar types of term.  For $\ell =
\tilde{\ell}_{cr}$, we recover the pure UV result (proportional to
$\ell/\epsilon$) and also the constant  $- R_{\rm UV}^2$, the
constant ensured by the fact that the disc preserves some conformal
invariance, as expected.  For large $\ell$, we have:
\begin{equation}\label{final_expand_4_disc}
\frac{\mathrm{Area}}{2 \pi} =  R_{\rm UV}^2 \frac{\ell}{\epsilon}  -  R_{\rm UV}^2 \frac{\ell}{\tilde{\ell}_{cr}} + R_{\rm IR}^2 \frac{\ell}{\tilde{\epsilon}}  - R_{\rm IR}^2 + O(\epsilon, 1/\ell)\ .
\end{equation}
So in the AdS$_4$ case, the constant term shifts from its UV result to
its IR result $-R_{\rm IR}^2$.    Again, in addition to the usual UV area law  (proportional to
$\ell/\epsilon$),  we have a new area law controlled by the domain wall:
\begin{equation}\label{eqt:newarealaw4}
  \ell\left( \frac{R_{\rm IR}^2}{\tilde{\epsilon}}-   \frac{R_{\rm UV}^2}{\tilde{\ell}_{cr}}\right)\ ,
\end{equation} 
which should be compared to the example from AdS$_5$ in
equation~\reef{eqt:new_area_law}. The same comments we made for the
new area law there apply here: It is not necessarily positive, but we
expect it to get more positive as the domain wall is sent to the UV,
where generically it gets thinner.

\subsection{The Case of AdS$_3$.}

Next we consider the case of AdS$_3$, pertaining to flows in $1+1$
dimensions. We use a spatial interval of length $2\ell$ for our region
${\cal A}$.  The area is given by:
\begin{eqnarray}
\frac{\mathrm{Area}}{2} &=& - R_{\rm UV} \ln \left( \frac{\epsilon}{\ell} \right) + R_{\rm UV} \ln(2) - R_{\rm UV} \tanh^{-1} \left( \frac{\sqrt{ \ell^2 - R_{\rm UV}^2 e^{-2 \frac{r_{DW}}{R_{\rm UV}}}}}{\ell} \right) \nonumber \\
&& + R_{\rm IR} \tanh^{-1} \left( \sqrt{ \frac{\ell^2 - R_{\rm UV}^2 e^{-2 \frac{r_{DW}}{R_{\rm UV}}}}{ \ell^2 - R_{\rm UV}^2 e^{-2 \frac{r_{DW}}{R_{\rm UV}}} + R_{\rm IR}^2 e^{-2 \frac{r_{DW}}{R_{\rm IR}}} }  } \right)  + O(\epsilon)
\end{eqnarray}
In the large $\ell$ limit, this gives
\begin{equation}
\frac{\mathrm{Area}}{2} =  - R_{\rm UV} \ln \left( \frac{\epsilon}{\tilde{\ell}_{cr}} \right) - R_{\rm UV} - R_{\rm IR} \ln \left( \frac{\tilde{\epsilon}}{\ell} \right) + R_{\rm IR} \ln(2) + O(\epsilon, 1/\ell) \ ,
\end{equation}
where $\tilde{\ell}_{cr}$ and $\tilde{\epsilon}$ are defined in equations \reef{eqt:tildeellcr} and \reef{eqt:tildeell} respectively.  So again we see that the universal coefficient (in front of the
natural logarithm) becomes the IR factor in the large $\ell$ limit.
The IR cutoff replaces the UV cutoff just as observed before.

\subsection{The Strip and AdS$_4$.}

We next consider an area $\cal A$ that is a strip in AdS$_4$, to
compare our results for the disc.  We take the strip to be of finite
width $\ell$ in the $x$ direction, and of length $L$ in the remaining
direction, which will be taken to be large, making an infinite
strip. The area is given by:
\begin{equation}
\mathrm{Area} = 2 L \int_{0}^{\ell/2} d x \ e^{2 A(r)} \sqrt{ 1 + e^{-2 A(r)} r'(x)^2 }\ .
\end{equation}
Since there is no explicit dependence on $x$, there is a constant of
motion in the dynamical problem associated to minimizing the area.
However, we must be careful since the constant of motion on either
side of the domain wall is not the same:
\begin{equation}
\frac{e^{2 A(r)}}{\sqrt{1 + e^{-2A(r)} r'(x)^2}} = \left\{ \begin{array}{lr}
e^{2 \frac{r_{\ast}}{R_{\rm UV}}} \ , & r > r_{DW} \\
e^{2 \frac{r_{\ast}}{R_{\rm IR}}} \ , & r < r_{DW} 
\end{array} \right. \ .
\end{equation}
On the IR side, the constant is simply given by $r'(x) = 0$, which
occurs at a radial position we will denote as $r_\ast$.  The constant
on the UV side is determined by asking that $r'(x) = 0$ when $r_\ast =
r_{DW}$, which is the critical situation before our embedding enters
the IR AdS.  We can in turn calculate the area and length in terms
of~$r_\ast$ and expand for small $\epsilon$:
\begin{eqnarray}
\frac{\mathrm{Area}}{2 L} &=& \frac{R_{\rm UV}^2}{\epsilon} - R_{\rm UV} e^{\frac{r_{DW}}{R_{\rm UV}}} \sqrt{ 1 - e^{4 \frac{r_\ast - r_{DW}}{R_{\rm UV}}}}  + R_{\rm IR} e^{\frac{r_{DW}}{R_{\rm IR}}} \sqrt{ 1 - e^{4 \frac{r_\ast - r_{DW}}{R_{\rm IR}}}} \nonumber \\
&& - e^{\frac{r_\ast}{R_{\rm IR}}} \frac{ \sqrt{\pi} R_{\rm IR} \Gamma \left(\frac{7}{4} \right)}{3 \Gamma \left( \frac{5}{4} \right)} - \frac{1}{3} e^{-3 \frac{r_{DW}}{R_{\rm UV}}+ 4 \frac{r_\ast}{R_{\rm UV}}} R_{\rm UV} {}_2 F_1 \left( \frac{1}{2}, \frac{3}{4}, \frac{7}{4}, e^{4 \frac{r_\ast - r_{DW}}{R_{\rm UV}}} \right) \nonumber \\
&& + \frac{1}{3} e^{-3 \frac{r_{DW}}{R_{\rm IR}}+ 4 \frac{r_\ast}{R_{\rm IR}}} R_{\rm IR} {}_2 F_1 \left( \frac{1}{2}, \frac{3}{4}, \frac{7}{4}, e^{4 \frac{r_\ast - r_{DW}}{R_{\rm IR}}} \right) \ .
\end{eqnarray}
\begin{eqnarray}
\frac{\ell}{2} &=&   e^{-\frac{r_\ast}{R_{\rm IR}}} \frac{ \sqrt{\pi} R_{\rm IR} \Gamma \left(\frac{7}{4} \right)}{3 \Gamma \left( \frac{5}{4} \right)} + \frac{1}{3} e^{- 3 \frac{r_{DW}}{R_{\rm UV}}+ 2 \frac{r_\ast}{R_{\rm UV}}} R_{\rm UV} {}_2 F_1 \left( \frac{1}{2}, \frac{3}{4}, \frac{7}{4}, e^{4 \frac{r_\ast - r_{DW}}{R_{\rm UV}}} \right) \nonumber \\
&& - \frac{1}{3} e^{-3 \frac{r_{DW}}{R_{\rm IR}}+ 2 \frac{r_\ast}{R_{\rm IR}}} R_{\rm IR} {}_2 F_1 \left( \frac{1}{2}, \frac{3}{4}, \frac{7}{4}, e^{4 \frac{r_\ast - r_{DW}}{R_{\rm IR}}} \right) \ .
\end{eqnarray}
Here, we see the appearance of the Gauss hypergeometric
function:\begin{equation}
  {}_2F_1(a,b,c;z)\equiv\frac{\Gamma(c)}{\Gamma(b)\Gamma(c-b)}\int_0^1\frac{t^{b-1}(1-t)^{c-b-1}}{(1-tz)^a}
  \,  dt\ .\end{equation}
The large $\ell$ limit corresponds to taking $r_\ast \to - \infty$, which gives us:
\begin{eqnarray}\label{eqt:newarealaw4_b}
\frac{\mathrm{Area}}{2 L} &=& \frac{R_{\rm UV}^2}{\epsilon}  - R_{\rm UV} e^{\frac{r_{DW}}{R_{\rm UV}}}   + R_{\rm IR} e^{\frac{r_{DW}}{R_{\rm IR}}} + O(\epsilon, -1/r_{\ast})\ .
\end{eqnarray}
So we see that that the constant term here is exactly the new area
law's coefficient that we saw in the disc case, in equation~\reef{eqt:newarealaw4}. Again, far enough in the UV, for large enough mass, our analysis suggests that this coefficient is  positive.

\subsection{The Box and AdS$_5$.}
Returning to AdS$_5$, we consider for region $\cal A$ a box in
AdS$_5$, in order to compare to the round ball we studied before.
Here the finite width is again $\ell$ and the two other sides are of
length~$L$, which we again take to be large. The computation proceeds
in a similar way.  The area gives:
\begin{eqnarray}
\frac{\mathrm{Area}}{2 L^2} &=& \frac{R_{\rm UV}^3}{2 \epsilon^2} - \frac{1}{2} R_{\rm UV} e^{2 \frac{r_{DW}}{R_{\rm UV}}} \sqrt{ 1 - e^{6 \frac{r_\ast - r_{DW}}{R_{\rm UV}}}}  + \frac{1}{2} R_{\rm IR} e^{2 \frac{r_{DW}}{R_{\rm IR}}} \sqrt{ 1 - e^{6 \frac{r_\ast - r_{DW}}{R_{\rm IR}}}} \nonumber \\
&& - e^{2 \frac{r_\ast}{R_{\rm IR}}} \frac{ \sqrt{\pi} R_{\rm IR} \Gamma \left(\frac{5}{3} \right)}{8 \Gamma \left( \frac{7}{6} \right)} - \frac{1}{8} e^{-4 \frac{r_{DW}}{R_{\rm UV}}+ 6 \frac{r_\ast}{R_{\rm UV}}} R_{\rm UV} {}_2 F_1 \left( \frac{1}{2}, \frac{2}{3}, \frac{5}{3}, e^{6 \frac{r_\ast - r_{DW}}{R_{\rm UV}}} \right) \nonumber  \\
&& + \frac{1}{8} e^{-4 \frac{r_{DW}}{R_{\rm IR}}+ 6 \frac{r_\ast}{R_{\rm IR}}} R_{\rm IR} {}_2 F_1 \left( \frac{1}{2}, \frac{2}{3}, \frac{5}{3}, e^{6 \frac{r_\ast - r_{DW}}{R_{\rm IR}}} \right) \ ,
\end{eqnarray}
and taking the $r_\ast \to -\infty$ limit gives:
\begin{eqnarray}
\frac{\mathrm{Area}}{2 L^2} &=& \frac{R_{\rm UV}^3}{2 \epsilon^2} - \frac{1}{2} R_{\rm UV} e^{2 \frac{r_{DW}}{R_{\rm UV}}}   + \frac{1}{2} R_{\rm IR} e^{2 \frac{r_{DW}}{R_{\rm IR}}} + O(\epsilon, -1/r_{\ast})\ .
\end{eqnarray}
Again, we have that the constant term has the same coefficient as the
new area law term for the ball case, as seen in
equation~\reef{eqt:new_area_law}.

\section{The  Four Dimensional Holographic RG Flow}


\subsection{The Holographic Dual Gravity Background}
In field theory terms, the RG flow is defined by an $\mathcal{N} = 1$
supersymmetric deformation of the $\mathcal{N}=4$ supersymmetric Yang
Mills theory given by introducing a mass term for one of the chiral
multiplets.
This relevant deformation causes the $\mathcal{N} = 4$ theory to flow
to an $\mathcal{N} = 1$ fixed point in the IR called the
Leigh--Strassler fixed
point~\cite{Karch:1999pv,Leigh:1995ep,Argyres:1999xu}. For the $SU(N)$
theory at large $N$, there is an holographic dual of this
physics~\cite{Khavaev:1998fb}, represented by a flow between two five
dimensional anti--de Sitter (AdS$_5$) fixed points of ${\cal N}=8$
gauged supergravity in five dimensions. One point has the maximal
$SO(6)$ symmetry, and the other has $SU(2)\times U(1)$, global
symmetries of the dual field theories.
The relevant five dimensional gauged supergravity action is
\cite{Khavaev:1998fb,Freedman:1999gp,Pilch:2000fu}:
\begin{equation}
S = \frac{1}{16 \pi G_5} \int d^5 x \sqrt{-g} \left( R - 2 \left( \partial \chi \right)^2 - 12 \left( \partial \alpha \right)^2 - 4 \mathcal{P} \right)\ ,
\end{equation}
with
\begin{equation}
\mathcal{P} = \frac{1}{2 R^2} \left( \frac{1}{6} \left( \frac{\partial W}{\partial \alpha} \right)^2 + \left( \frac{\partial W}{\partial \chi} \right)^2 \right) - \frac{4}{3 R^2} W^2\ ,
\end{equation}
where the superpotential $W$ is given by:
\begin{equation}
W = \frac{1}{4 \rho^2} \left( \cosh(2 \chi) \left( \rho^6 - 2 \right) - \left( 3 \rho^6 + 2 \right) \right) \ ,
\end{equation}
with $\rho = \exp (\alpha) $.  The scalar field $\chi$ is dual to an
operator of dimension three in the field theory while the scalar field
$\alpha$ is dual to a dimension two operator:
\begin{equation}
\alpha: \sum_{i = 1}^4 \mathrm{Tr} \left( \phi_i \phi_i \right) - 2 \sum_{i = 5}^6 \mathrm{Tr} \left( \phi_i \phi_i \right) \ , \quad \chi: \mathrm{Tr} \left( \lambda_3 \lambda_3 + \varphi_1 \left[ \varphi_2 , \varphi_3 \right] \right) + \mathrm{h.c.}\ ,
\end{equation}
where $\varphi_k = \phi_{2k - 1} + i \phi_{2 k} $, $k=1,\hdots,
3$. Here $\phi_i$ ($i=1,\hdots, 6$) are the six scalars in the ${\cal
  N}=4$ multiplet, and the $\lambda_k$ (three of that adjoint
multiplet's four fermions) are ${\cal N}=1$ partners of the
$\varphi_k$, forming the three chiral multiplets.  This combination of
operators is exactly what is needed to reproduce the deformation.
The geometry in five dimensions, of domain wall form, can be
parametrised in the following manner:
\begin{equation}
ds_{1,4}^2 = e^{2 A(r)} \left( -dt^2 + dx_1^2 + dx_2^2 + dx_3^2 \right) + dr^2 \ .
\end{equation}
The supergravity equations of motion yield the following flow
equations:
\begin{eqnarray}
\frac{d \alpha}{d r} &=& \frac{e^\alpha}{6 R}  \frac{\partial W}{\partial \alpha} = \frac{1}{6 R} \left(\frac{e^{6 \alpha} \left( \cosh(2 \chi) - 3 \right) + \cosh(2 \chi) +1}{e^{2 \alpha} } \right)\ , \nonumber \\
\frac{d \chi}{d r} &=& \frac{1}{R} \frac{\partial W}{\partial \chi} = \frac{1}{2R} \left( \frac{ \left( e^{6 \alpha}  - 2 \right) \sinh(2 \chi)}{e^{2 \alpha} } \right)\ , \nonumber\\
\frac{d A}{d r} &=& - \frac{2}{3 R} W =- \frac{1}{6 R} \frac{\cosh(2 \chi)\left( e^{6 \alpha}  -2 \right) - \left( 3 e^{6 \alpha}  +2 \right)}{e^{2 \alpha} }\ .
\end{eqnarray}
In these coordinates, the UV is at $r\to+\infty$ and the IR is at
$r\to\-\infty$, as in earlier sections. In either limit, the right
hand side of the first two equations vanish, and the scalars run to
specific values ($\alpha=0, \chi=0$ at one end, $\alpha=\frac16\ln2,
\chi=\frac12\ln3$ at the other), while $A(r)$ becomes $\frac{r}{R}$ in
the UV and $\frac{2^{5/3}}{3} \frac{r} {R}$ in the IR, defining an
AdS$_5$ in each case, and hence a conformally invariant dual field
theory at each end. This is a fat, smooth version of our simple thin
domain wall model of the previous sections. Here $R_{\rm UV}=R$ and
$R_{\rm IR}=3R/2^{5/3}$.

To study the UV behavior of the fields, we find it convenient to
define a coordinate $\tilde{z}$ given by:
\begin{equation} \label{eqt:tildez}
\tilde{z} =  e^{- r/ R} \ , 
\end{equation}
and we find the asymptotic behavior of the fields near $\tilde{z} = 0$ (UV AdS boundary):
\begin{eqnarray}
\chi(\tilde{z}) &=& \tilde{z} \left( a_0 + \tilde{z}^2 \left( -a_0^3 -4 a_0 a_1 + \ln(\tilde{z}) \frac{8}{3} a_0^3 \right) + O(\tilde{z}^4) \right)\ , \nonumber \\
\alpha(\tilde{z}) &=&  \tilde{z}^2 \left( a_1 + \ln(\tilde{z}) \left( - \frac{2}{3} a_0^2 \right) \right)+ O(\tilde{z}^4)\ , \nonumber \\
A(\tilde{z}) &=& - \ln(\tilde{z}) + A_0  - \frac{1}{6} a_0^2 \tilde{z}^2  + O(\tilde{z}^4)\ . \label{eqt:A(z)}
\end{eqnarray}
The constant $a_0$ is related to the mass of the $\Phi_3$ multiplet
\emph{via} \cite{Johnson:2001ze}:
\begin{equation}\label{eqn:mass_three}
m_3 = \frac{2 a_0}{R}\ .
\end{equation}
To study the IR behavior of the fields, we define a coordinate $\tilde{u}$ given by:
\begin{equation}
\tilde{u} = e^{\lambda r / R} \ ,
\end{equation}
where $\lambda = \frac{2^{5/3} \left( \sqrt{7}-1 \right)}{3}$.  The
asymptotic (near ${\tilde u} = 0$) behavior in the IR is given by:
\begin{eqnarray} \label{eqt:IRBC}
\chi(\tilde{u}) &=& \frac{1}{2} \ln(3) + \tilde{u} b_0 + O(\tilde{u}^2) \label{eqt:IRchi}\ ,  \nonumber \\
\alpha(\tilde{u}) &=& \frac{1}{6} \ln(2) + \tilde{u} \left( \frac{\sqrt{7}-1}{6} \right) b_0  + O(\tilde{u}^2) \label{eqt:IRrho}\ , \nonumber \\
A(\tilde{u}) &=& \frac{1}{\sqrt{7}-1} \ln(\tilde{u}) + B_0 + O(\tilde{u}^2) \label{eqt:IRA}\ .
\end{eqnarray}

\subsection{Numerically Solving for the Flow}
%
To solve the flow numerically (as we will need to do in order to
compute the entanglement entropy), it is convenient to work with a
coordinate:
\begin{equation}
  x = \tilde{z}^2\ ,
\end{equation}
and employ a shooting method to solve the equations. To shoot from the
IR we take $x_{max} = 10^6$, and towards the UV we take $x =
\epsilon$, where we use $\epsilon = 10^{-12}$.  To get good numerical
stability, we define new fields $(\beta, \eta)$,
\begin{equation}
\alpha(x) = x \beta(x) \ , \quad \chi(x) = x^{1/2} \eta(x) \ , \quad A(x) = - \frac{1}{2} \ln(x) + a(x)\ ,
\end{equation}
such that near the AdS boundary, the leading behavior of these fields is given by:
\begin{equation}
\eta(x) = a_0 + O(x) \ , \quad \beta(x) = a_1 - \frac{1}{3} a_0^2 \ln(x) + O(x) \ , \quad a(x) = A_0 - \frac{1}{6} a_0^2 x + O(x^2) \ .
\end{equation}
In the IR, we use the results in equation \reef{eqt:IRBC} (up to
$O({\tilde u}^4)$) as our shooting conditions (we have the freedom of
choosing the parameter $b_1$, which is always less than zero).  We can
then extract the values of $a_0$ and~$a_1$ at the AdS boundary for our
solution.  Note that the choice of $B_0$ will determine the choice of
$A_0$.
We can choose to eliminate the constant $A_0$ by appropriately
rescaling our coordinate $\tilde{z}$.  In particular, if we solve our
equations with $B_0 = 0$, we can extract the constant $A_0$, and then
simply perform the following transformation to eliminate it from our
metric:
 \begin{equation}
 \tilde{z} \to e^{A_0} \tilde{z}\ .
 \end{equation}
 Finally, as indicated in ref.~\cite{Freedman:1999gp}, there are
 constants of the motion regardless of the choice of $b_1$ (and
 subsequently, the choice of $(a_0, a_1)$), given by:
\begin{equation}
-b_1 a_0^\lambda \approx 0.1493 \ , \quad \frac{\sqrt{6} a_1}{a_0^2} + \sqrt{\frac{8}{3}} \ln(a_0) \approx - 1.4696 \ .
\end{equation}
These constants, for a given flow, set the size and position of the
domain wall separating the two AdS$_5$ regions of the geometry.

\begin{figure}[ht]
\begin{center}
\subfigure[]{\includegraphics[width=3in]{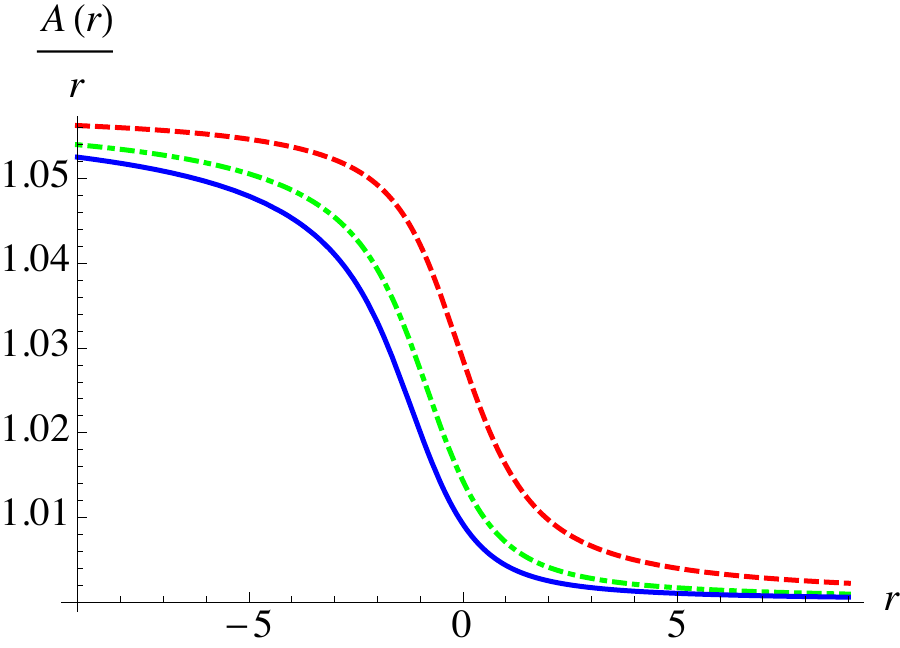}\label{fig:wall1}} \hspace{0.5cm}
\subfigure[]{\includegraphics[width=3in]{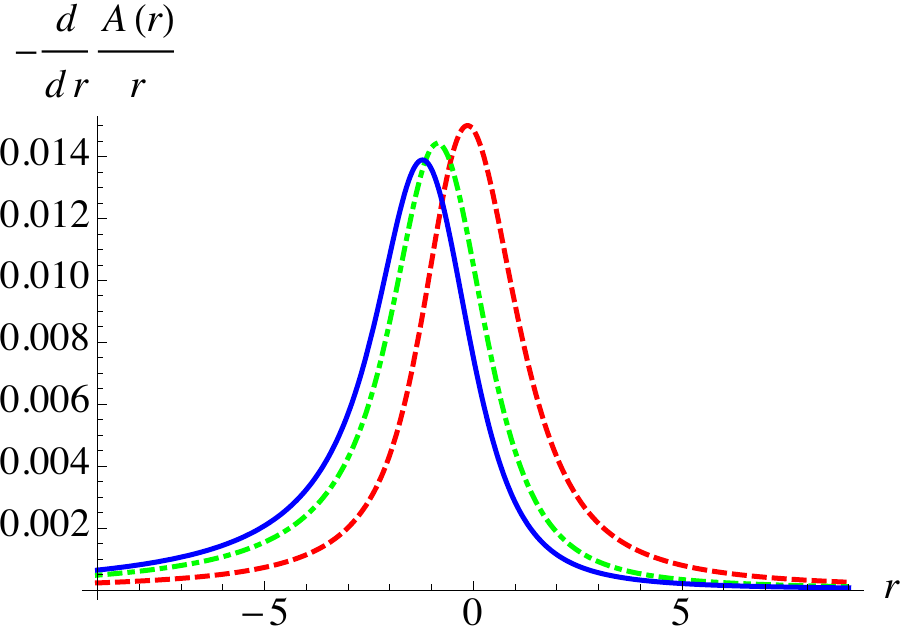}\label{fig:wall2}} 
\caption{\small Samples of the domain wall behaviour. The lowest (blue
  solid) curve corresponds to $b_1 \to - 3$), next lowest (green
  dot--dashed) curve is with $b_1 = -2$, and the top (red dashed
  curve) is with $b_1 = - 1$. The UV is to the right, and the mass
  increases with increasing (toward the positive) $b_1$.}
   \label{fig:wall}
   \end{center}
   \end{figure}

It is instructive to look at the results to get a sense of what a real
domain wall looks like in these solutions. We plot (in the original
radial coordinate $r$) the metric function $A(r)/r$, and also
$-A(r)^\prime$ in figure~\ref{fig:wall}.
   The wall noticeably gets thinner as it moves toward the UV. While
   it would be instructive to track the wall toward the UV for very
   large values of the mass, it becomes harder to control the
   numerical accuracy in that regime.


\subsection{The Holographic Entanglement Entropy}
%

\subsubsection{The Box}
%
The metric we need is given by:
\begin{equation}
ds_{1,4}^2 = \frac{R^2}{z^2} e^{2 \left(a(\tilde{z})- a(0) \right)} \left( - dt^2 + d x_1^2 + dx_2^2 + dx_3^2 \right) + \frac{R^2}{z^2} dz^2\ .
\end{equation}
It is important to notice that we are working here in a rescaled $z$ coordinate.  $\tilde{z}$ is the coordinate of the previous section, with $\tilde{z} = z \exp(a(0)) / R$. This will mean that the mass of the chiral multiplet (controlled by the scalar $\chi$; see equation~\reef{eqn:mass_three}) will now not be set by $a_0$, but by $e^{a(0)}a_0$.  

We consider a strip of infinite extent in the $(x_2, x_3)$ directions and finite extent in the $x_1$ direction.  We take as coordinates and ansatz for our surface embedding:
\begin{equation}
\xi_1 = x_1 \ , \quad \xi_2 = x_2 \ , \quad \xi_3 = x_3 \ , \quad z \equiv z(x_1)\ .
\end{equation}
The area of the surface is then given by:
\begin{equation}
\mathrm{Area} = R^3 L^2 \int_{- \ell/2}^{\ell/2} dx_1 \frac{e^{3 \left(a(\tilde{z}) - a(0) \right)}}{z^3} \left( 1 + e^{-2 \left(a(\tilde{z}) - a(0) \right)} z'(x_1)^2 \right)^{1/2} \ .
\end{equation}
Since there is no explicit dependence on the coordinate $x_1$, a constant of the ``motion'' is:
\begin{equation}
P = \mathcal{L} - z'(x_1) \frac{\partial \mathcal{L}}{\partial z'(x_1)} = \frac{e^{3 \left(a(\tilde{z}) - a(0) \right)}}{z(x_1)^3 \sqrt{ 1+ e^{-2 \left(a(\tilde{z}) - a(0) \right)} z'(x_1)^2)}}\ .
\end{equation}
We can evaluate this constant at the turning point $x_1 = 0$ (with corresponding value $z(0) = z_\ast$) and rewrite our area integral in terms of $z$:
\begin{equation} \label{eqt:Area}
\mathrm{Area} = 2 R^3 L^2 \int_{\epsilon}^{z_\ast} d z \frac{e^{2 (a(\tilde{z}) - a(0))}}{z^3 \sqrt{1- e^{-6(a(\tilde{z}) - a_\ast)} \frac{z^6}{z_\ast^6} }}\ ,
\end{equation}
where we have defined our UV cutoff $\epsilon$ as $z(\pm \ell/2) = \epsilon$.  Given the solution for the background, this expression for the area can be simply integrated numerically.  Furthermore, we can write $z_\ast$ in terms of $\ell$ as:
\begin{equation}
\frac{\ell}{2} = \int^{z_\ast}_\epsilon dz  \frac{ e^{- \left(a(\tilde{z}) - a(0) \right)} }{\sqrt{ e^{6 \left( a(\tilde{z}) - a_\ast \right)} \frac{z_\ast^6}{z^6} -1}}\ .
\end{equation}
For the case of pure AdS (which simply corresponds to taking $a(\tilde{z}) = a_\ast =  a(0)$ in our equations), equation~\reef{eqt:Area} has a single divergent term near the AdS boundary proportional to $\epsilon^{-2}$.
For the flow geometry, the non--trivial behavior of $a(\tilde{z})$ produces a new divergence in addition to the pure AdS one.  Expanding the exponential in the numerator, we find that near the AdS boundary, the divergent terms produced by the integral are:
\begin{equation} \label{eqt:UV_div}
\mathrm{Area}_{\rm UV} = 2 R^3 L^2 \int_{\epsilon} \frac{ 1 - \frac{a_0^2}{3 R^2} e^{2 a(0)} z^2}{ z^3} = R^3 L^2 \left(\frac{1}{\epsilon^2}  + \frac{2 a_0^2}{3 R^2} e^{2 a(0)} \ln\left( \frac{\epsilon}{R} \right) \right)\ .
\end{equation}
The $\ln(\epsilon)$ divergence is new.  Such new divergence terms associated with the function $a(z)$ are expected in $d \geq 4$.

\begin{figure}[ht] 
   \centering
   \includegraphics[width=3in]{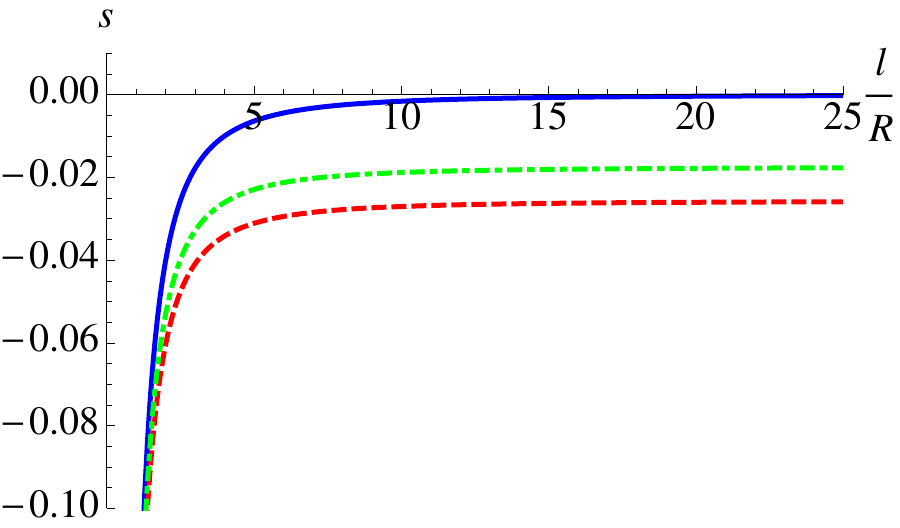} 
   \caption{\small Blue solid curve is pure AdS$_5$ result (corresponds to $b_1 \to - \infty$), green dot--dashed curve is with $b_1 = -2$, and the red dashed curve is with $b_1 =  - 1$}
   \label{fig:5dstrip}
\end{figure}

\begin{figure}[ht] 
   \centering
   \includegraphics[width=3in]{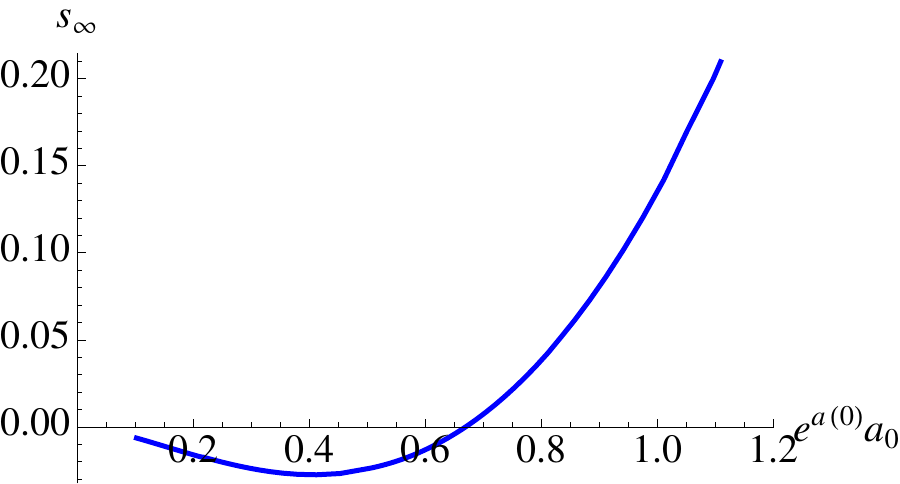} 
   \caption{\small Value of the asymptotic ($\ell/R \to \infty$) entanglement entropy for changing multiplet mass.  Recall that the multiplet mass is proportional to $e^{a(0)} a_0$.}
   \label{fig:strip_s_vs_b}
\end{figure}

\begin{figure}[ht]
\begin{center}
\subfigure[]{\includegraphics[width=3in]{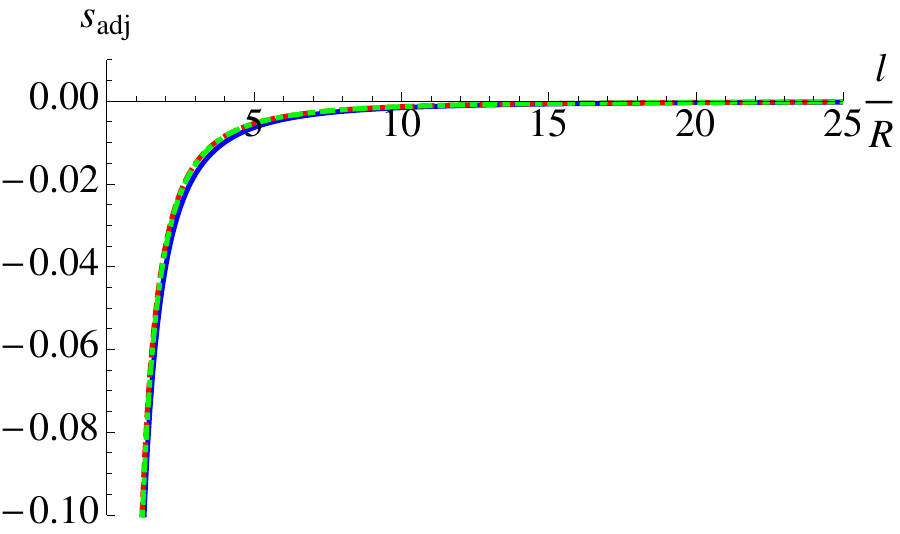}\label{fig:a}} \hspace{0.5cm}
\subfigure[]{\includegraphics[width=3in]{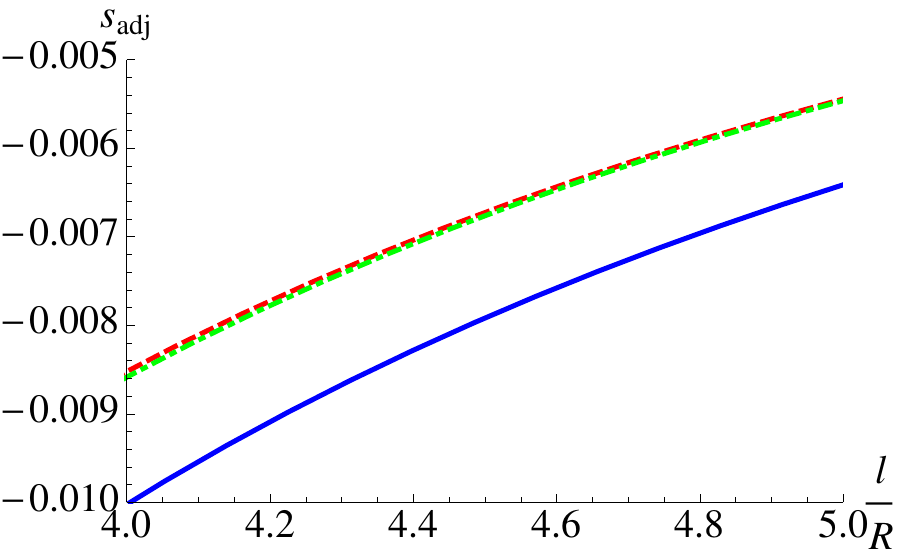}\label{fig:b}}
   \caption{\small The adjusted entanglement entropy.  Blue solid curve is pure AdS$_5$ result (corresponds to $b_1 \to - \infty$), green dot--dashed curve is with $b_1 = -2$, and the red dashed curve is with $b_1 =  - 1$.  The curves do not perfectly overlap when inspected at a higher magnification. } 
   \label{fig:Adjusted_strip}
   \end{center}
   \end{figure}

We will only be interested in the finite contribution (beyond this UV divergent term) to the entanglement entropy, and we will denote this finite contribution by $s$ such that:
\begin{equation} \label{eqt:little_s}
{4 G_{N}^{(5)}}S = {\mathrm{Area}} =  \left( \frac{R^3 L^2}{\epsilon^2} + \frac{2 a_0^2 R L^2 }{3} e^{2 a(0)} \ln\left( \frac{\epsilon}{R} \right)  + 2 R L^2 s \right)\ .
\end{equation}
To perform these integrals numerically, we find that using $\epsilon = 10^{-5}$ and using the coordinate $y = z^3$ gives reliable results.  We present some results in figure \ref{fig:5dstrip}.  
%
%
The key result is that we find that the entanglement entropy for the flow geometry asymptotes (for large $\ell/R$) to a constant value not equal to zero, as was predicted by the sharp domain wall analysis.  The value of the asymptotic constant, which we denote by $s_\infty$, can be understood as stemming from the parts of the surface in the interpolating region from one AdS geometry to the other. With our normalization of $s$ as defined in equation \reef{eqt:little_s}, the  interpolating part of the surface contributes as:
\begin{equation} \label{eqt:drop}
s_\infty =  R^2 \int_\epsilon^{z_\ast} d z \frac{e^{2(a(\tilde{z})-a_0)}}{z^3} \ .
\end{equation}
Taking $z_\ast \to \infty$, we find that the finite contribution of this area  matches extremely well the finite value for the entanglement entropy curves in figure \ref{fig:5dstrip}.  We present the dependence of $s_\infty$ on the multiplet mass in figure \ref{fig:strip_s_vs_b}.  We see that for large mass, the value increases as predicted, while for small mass, it actually decreases.  We can understand this behavior as follows.  The sharp domain wall analysis of the previous section applies when the mass is large and correspondingly the domain wall's thickness is small.  For small mass, the wall is fat, and so we can expect deviations (in this case, very large) from our predictions.  As the mass increases, the sharp domain wall approximation is approached, and our results fit better.
Although we have a clear understanding of this asymptotic behavior from our sharp domain wall analysis, where we argued it mixes the UV and IR physics around the domain wall, in this example we can completely characterize the behaviour in terms of the chiral multiplet mass.  We define a function $\Omega$ as follows:
\begin{equation}\label{eqn:define_omega}
\Omega = \Omega_0 \left( e^{a(0)} a_0 \right)^{-1} \ , 
\end{equation}
where $\Omega_0 \approx 0.6592 R$ and is a constant which we found from fitting to the data.  Using this function, we write the entanglement entropy as:
\begin{equation} \label{eqt:little_s_tilde}
{4 G_N^{(5)}}S = {\mathrm{Area}} = \left( \frac{R^3 L^2}{\epsilon^2} + \frac{2 a_0^2 R L^2 }{3} e^{2 a(0)} \ln\left( \frac{\epsilon}{\Omega} \right)  + 2 R L^2 s_{\mathrm{adj}} \right)\ ,
\end{equation}
where $s_{\mathrm{adj}}$ now asymptotes to zero.  We show this in figure \ref{fig:Adjusted_strip}.  A nice feature of the function $\Omega$ is that it captures the IR/UV domain wall physics in terms of a single length scale which naturally normalizes the new UV divergence.  This length scale emerges naturally from our holographic calculation, and it would be difficult (yet worthwhile) to independently calculate it in the dual field theory.  Furthermore the non--UV divergent parts of the entropy now  asymptotes to zero for large $\ell$, as appropriate for the entanglement entropy for the box region in this limit since there will only be the region ${\cal A}$ remaining.
\subsubsection{The Ball}
%
The metric we need is given by:
\begin{equation}
ds_{1,4}^2 = \frac{R^2}{z^2} e^{2 \left(a(z) - a(0) \right)} \left( - dt^2 + d r^2 + r^2 d \Omega_2^2 \right) + \frac{R^2}{z^2} dz^2\ .
\end{equation}
We consider a  three dimensional ball with radius $\ell$, so take as coordinates and ansatz for our surface embedding:
\begin{equation}
\xi_1 = r \ , \quad \xi_2 = \theta \ , \quad \xi_3 = \phi \ , \quad z \equiv z(r)\ .
\end{equation}
The area of the surface is then given by:
\begin{equation} \label{eqt:ball_area}
\mathrm{Area} = 4 \pi  R^3  \int_{0}^{\ell} dr \frac{r^2  e^{3 \left(a(\tilde{z})- a(0) \right)}}{z(r)^3} \left( 1 + e^{-2 \left(a(\tilde{z})- a(0) \right)} z'(r)^2 \right)^{1/2} \ .
\end{equation}
Since there is an explicit $r$ dependence, there is no conserved quantity (as was the case for the strip), so the extremal area must be found by solving the equations of motion derived from the above action:
\begin{eqnarray}
&&z''(r) + \frac{2 e^{-2 \left(a(\tilde{z})- a(0) \right)}}{r} z'(r)^3 + \left( \frac{3}{z(r)} - 8 e^{2 a(0)} z(r) a'(\tilde{z}) \right) z'(r)^2  \nonumber \\ 
&&\hskip4.5cm+\frac{2}{r} z'(r) + \frac{3 e^{2 \left(a(\tilde{z})- a(0) \right)}( 1-2  e^{2a(0)}z(r)^2 a'(\tilde{z}))}{z(r)} = 0\ ,
\end{eqnarray}
with the boundary conditions:
\begin{equation}
z(0) = z_\ast \ , \quad z(\ell) = \epsilon\ .
\end{equation}
In order to study the UV divergences of equation \reef{eqt:ball_area}, it is convenient to define a new coordinate:
\begin{equation}
y(r) =\frac{1}{\ell} \sqrt{\ell^2 + \epsilon^2 - r^2}\ .
\end{equation}
The reason for this particular choice is that, in the pure AdS geometry, the embedding $z(r)$ has a solution given by $z(r) = \ell y(r)$.  In terms of $y$, the embedding has asymptotic solution given by:
\begin{equation}
z(y) = \ell y + z_2 y^3 - \frac{\ell^3}{6 R^2} e^{2 a(0)} a_0^2  y^3 \ln(y)+ \cdots \ ,
\end{equation}
where $z_2$ is a constant chosen such that the solution satisfies $z'(y = 1) = 0$.  

\begin{figure}[ht] 
   \centering
   \includegraphics[width=3in]{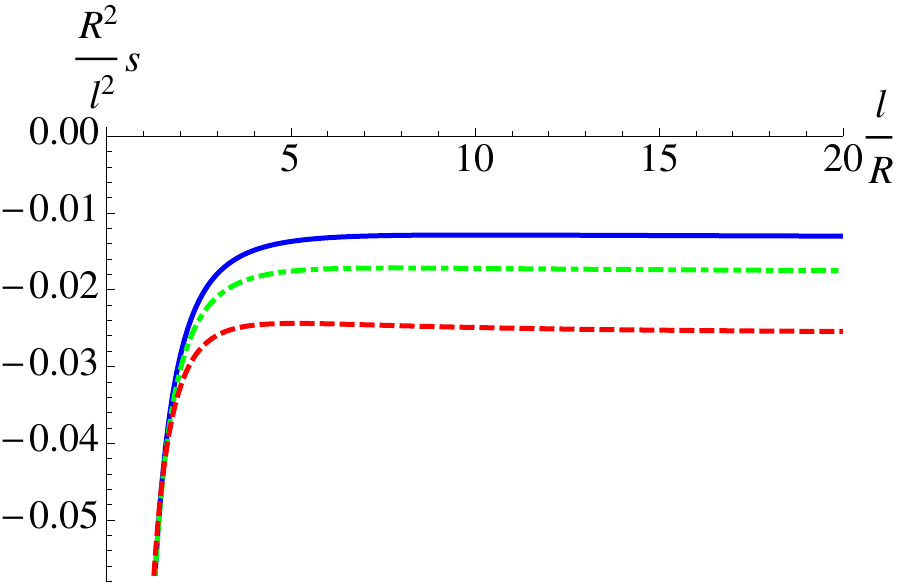} 
   \caption{\small Results for the ball with radius $\ell$.  Blue solid curve is for $-b_0 = 3$, green dot--dashed curve is for $-b_0 = 2$, and red dashed curve is for $-b_0 = 1$.}
   \label{fig:EEsphere}
\end{figure}

\begin{figure}[ht] 
   \centering
   \includegraphics[width=3in]{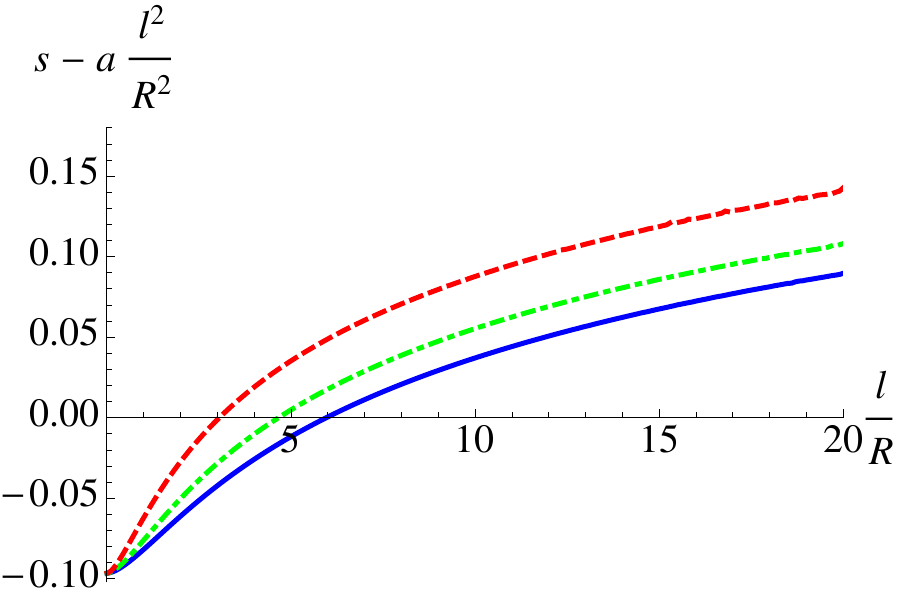} 
   \caption{\small  Entanglement entropy without the leading $\ell^2$ contribution.  Blue solid curve is for $-b_0 = 3$, green dot--dashed curve is for $-b_0 = 2$, and red dashed curve is for $-b_0 = 1$.}
   \label{fig:Adjusted_ball}
\end{figure}

\begin{figure}[ht]
\begin{center}
\subfigure[$a$ coefficient]{\includegraphics[width=3in]{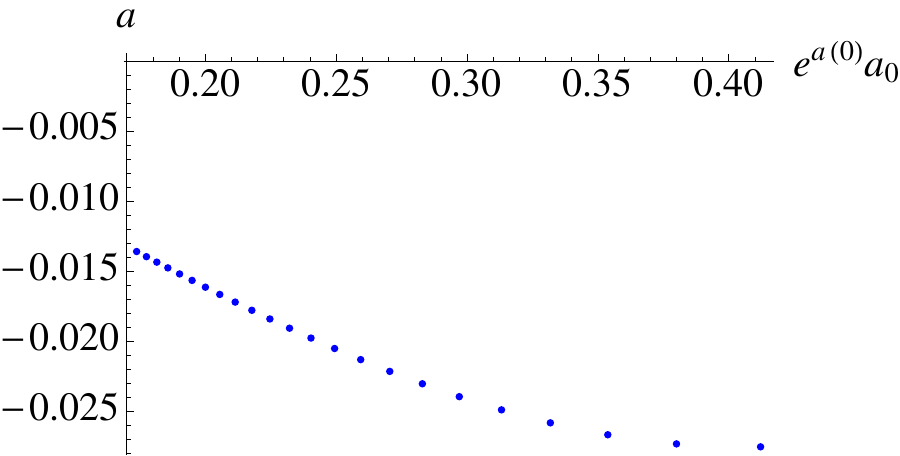}\label{fig:a_bit}} \hspace{0.5cm}
\subfigure[$b$ coefficient]{\includegraphics[width=3in]{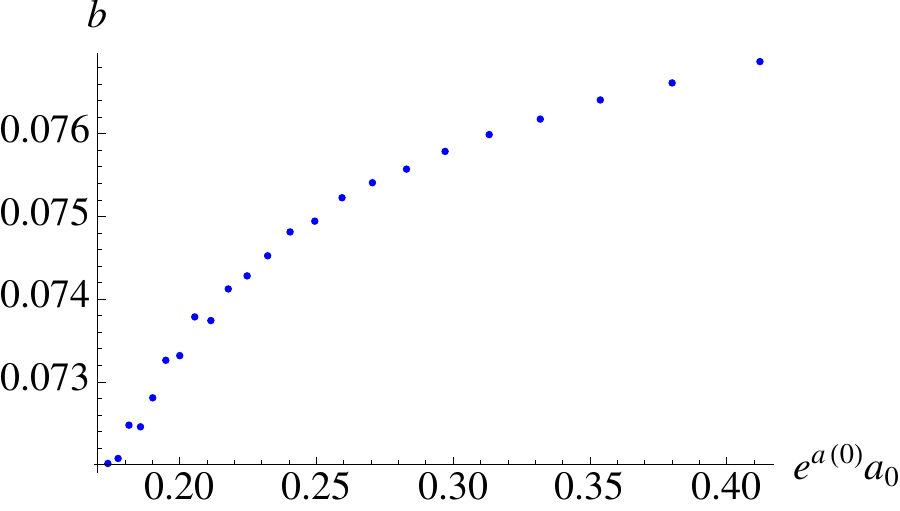}\label{fig:b_bit}} \hspace{0.5cm}
\subfigure[$c$ coefficient]{\includegraphics[width=3in]{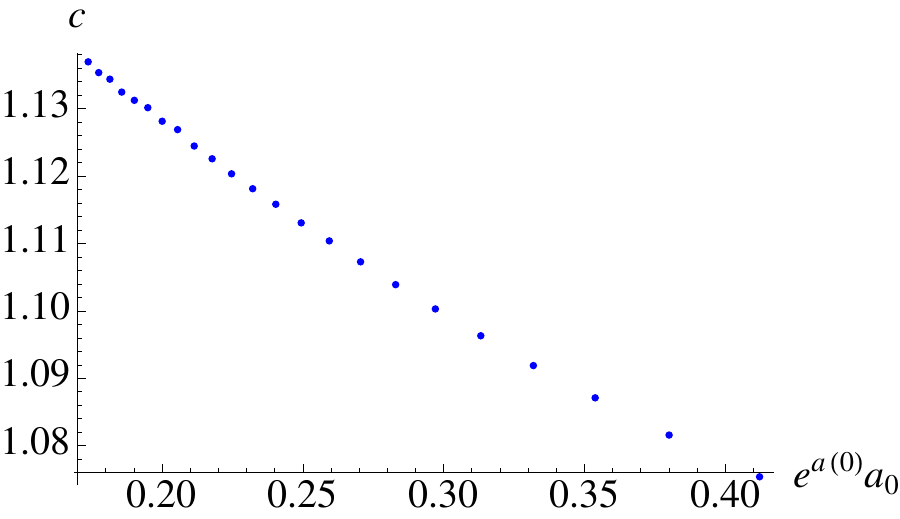}\label{fig:c_fit}} 
   \caption{\small Values of the fit coefficients of equation \reef{eqt:fit_ball}.}
   \label{fig:ball_mass_fit}
   \end{center}
   \end{figure}

\begin{figure}[ht] 
   \centering
   \includegraphics[width=3in]{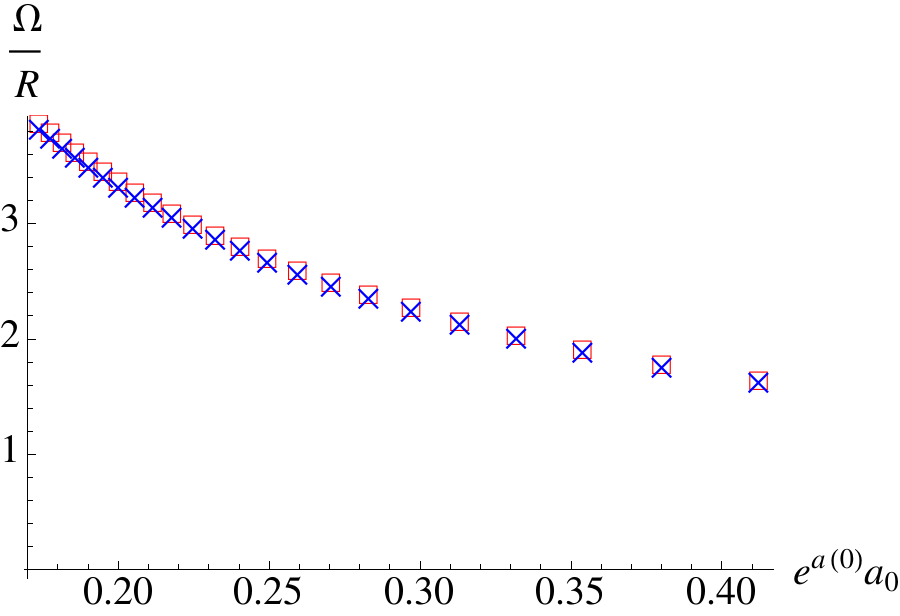} 
   \caption{\small  Comparing the prediction for the function $\Omega$ for the strip and the ball.  Blue crosses are the ball results and the red squares are the strip results.}
   \label{fig:Omega_Strip_vs_Ball}
\end{figure}

Let us define as before a quantity $s$ as follows:
\begin{equation}\label{eqn:shift_def_s}
\frac{\mathrm{Area}}{4 \pi} =  R^3 \left( \frac{\ell^2}{2 \epsilon^2} + \frac{1}{2} \ln \left( \frac{\epsilon}{\ell} \right) + \frac{\ell^2}{3 R^2} e^{2 a(0)} a_0^2 \ln \left( \frac{\epsilon}{R} \right)  + s\ . \right)
\end{equation} 
%
Our numerical results for $s$ are shown in figure \ref{fig:EEsphere}.  As expected from our sharp domain wall analysis, the asymptotic behavior is dominated by an $\ell^2$ behavior.  Furthermore, upon subtraction of this behavior, we find our expected $\ln(\ell)$ behavior (see figure \ref{fig:Adjusted_ball}).  In fact, the large $\ell$ behavior of the adjusted entanglement entropy is very well approximated by the expression:
\begin{equation} \label{eqt:fit_ball}
\lim_{\ell \to \infty} s  \approx a \left(\frac{\ell}{R} \right)^2 + b \ln \left(\frac{\ell}{R} \right) + c\ .
\end{equation}
Comparing to equation \reef{eqt:finalexpand5}, the coefficients $a$ and $c$ should be interpreted as being a complicated function of the domain wall data (and hence the mass term), whereas the coefficient $b$ should simply be related to the central charge of the IR theory.  We show the dependence of $a$, $b$, and $c$ on the mass in figure \ref{fig:ball_mass_fit}.
First, we see that the coefficient of the area term $a$ decreases with increasing mass, which deviates from our domain wall analysis.  However, our mass range displayed here is actually  in the same region where the analogous results for the box were decreasing. This is  all traceable (as it was then) to being far from the thin domain wall regime. In fact it is apparent that our results are just about to turn around, as happened for the box.    We may ask whether we can capture the behavior of this coefficient with a function $\Omega$, defined in a similar way to that which we did for the box (see equation~\reef{eqn:define_omega} and discussion below it). This turns out to be the case, and  we compare the function $\Omega$ for the strip and for the ball in figure \ref{fig:Omega_Strip_vs_Ball} ( where we again let the data determine the value $\Omega_0$) and we find that the results are remarkably close, suggesting that indeed this is a physical length scale in the theory. Since the only other scale in the theory is the multiplet mass, we expect that $\Omega$ is set by the inverse of the value of this mass after it has renormalized under RG flow to its value set by the scale at the domain wall.

\begin{figure}[ht] 
   \centering
   \includegraphics[width=3in]{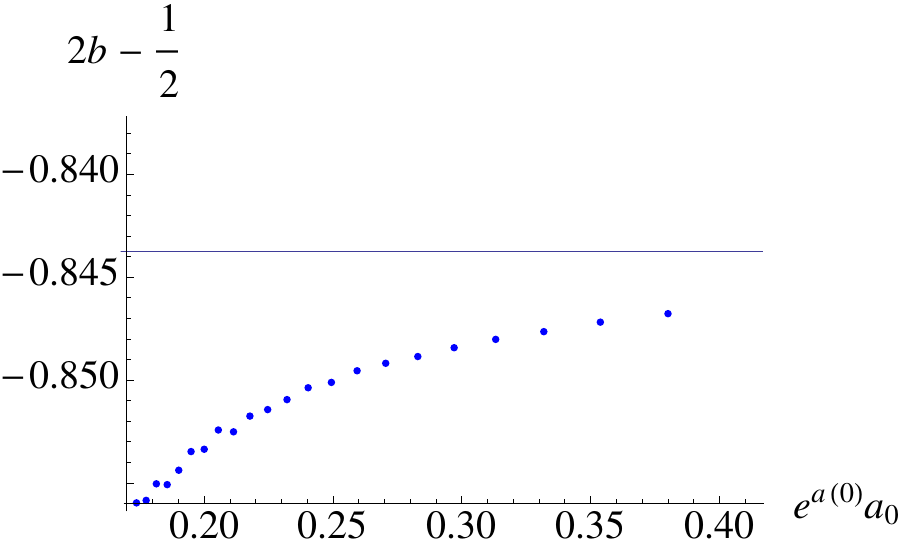} 
   \caption{\small  The coefficient of $\ln(\ell)$ for large $\ell$.  The solid line is the value of $-R_{\rm IR}^3/R^3$.}
   \label{fig:R_IR}
\end{figure}

Finally, it is interesting  to check whether the coefficient of the $\ln(\ell)$ term of the entanglement entropy indeed matches our expectation that for large enough $\ell$ it should be equal to $-R_{\rm IR}^3/2R^3$ (compare the definition of $s$ in equation~\reef{eqn:shift_def_s}, and $b$ in equation~\reef{eqt:fit_ball} and~\reef{eqt:finalexpand5}).  We present this check in figure~\ref{fig:R_IR}.  Indeed, we find that our results are very close to the expected value, and in fact the results get better with larger mass.  It is not entirely unexpected that the results deviate somewhat for smaller mass. With smaller masses, one needs to go to larger $\ell$ (and larger AdS radial position) to enter the IR AdS, and larger $\ell$ is increasingly hard to explore numerically.
%
\section{The Three Dimensional Holographic RG Flow}
%
\subsection{The Holographic Dual Gravity Background}
We consider the 2+1 flow of ref.~\cite{Corrado:2001nv}.  It is the analogue of the Leigh--Strassler flow we studied earlier.  The IR fixed point is the $\mathcal{N}=2$ fixed point studied in \cite{Warner:1983vz,Nicolai:1985hs}.  Starting with an ansatz for the metric of the form:
\begin{equation}
ds_{1,3}^2 = e^{2 A(r)} \left( - dt^2 + dx_1^2 + dx_2^2 \right) + dr^2\ ,
\end{equation}
the flow equations are given by\footnote{We use a different definition of the AdS radius $R$ than the authors of ref.~\cite{Corrado:2001nv}.  The reason for this is such that the asymptotic behavior in terms of the coordinate $r$ are the same for our previous analysis.}:
\begin{eqnarray}
A'(r) &=& - \frac{1}{R} W \ , \nonumber\\
\rho'(r) &=& \frac{1}{16 R} \frac{ \left( \cosh(2 \chi) + 1 \right) + \rho^8 \left( \cosh(2 \chi) - 3 \right)}{\rho} \ , \nonumber \\
\chi'(r) &=& \frac{1}{4 R} \frac{ \left( \rho^8 - 3 \right) \sinh(2 \chi) }{\rho^2 }\ .
\end{eqnarray}
To study the UV asymptotics, we define a coordinate $\tilde{z}$ via:
\begin{equation}
\tilde{z} = e^{- r /R}\ ,
\end{equation}
and find that the fields have the following asymptotics:
\begin{eqnarray}
A(\tilde{z}) &=& - \ln(\tilde{z}) - \frac{\tilde{z}^2}{4} \left(  12 a_1^2 + a_0^2 \right) + O(\tilde{z}^3)\ , \nonumber\\
\chi(\tilde{z}) &=& \tilde{z} a_0 - 6 a_1 a_0 \tilde{z}^2  + O(\tilde{z}^3)\ , \nonumber \\
\alpha(\tilde{z}) &=& \tilde{z} a_1 + \left( 2 a_1^2 - \frac{a_0^2}{4} \right)\tilde{z^2} + O(\tilde{z}^3)\ .
\end{eqnarray}
In the IR, we define a coordinate $\tilde{u}$ \emph{via}:
\begin{equation}
\tilde{u} = e^{\lambda r /R}\ , 
\end{equation}
where $\lambda = 3^{3/4} \left( \sqrt{17}-1 \right)/4$.  The asymptotics of the field is given by:
\begin{eqnarray}
A(\tilde{u}) &=& \frac{2}{\sqrt{17}-1} \ln(\tilde{u}) + B_0 + O(\tilde{u}^2) \ ,\nonumber\\
\chi(\tilde{u}) &=& \frac{1}{2} \cosh^{-1} (2) +  b_0 \tilde{u} + O(\tilde{u}^2) \ , \nonumber\\
\alpha(\tilde{u}) &=& \frac{1}{8} \ln \left(3 \right) + \frac{\sqrt{17}-1}{8 \sqrt{3}} b_0 \tilde{u} + O(\tilde{u}^2)\ .
\end{eqnarray}
Note that this gives that $R_{\rm IR} = 2 R / 3^{3/4}$.  
\subsection{The Entanglement Entropy}
\begin{figure}[ht]
\begin{center}
\subfigure[Strip]{\includegraphics[width=3in]{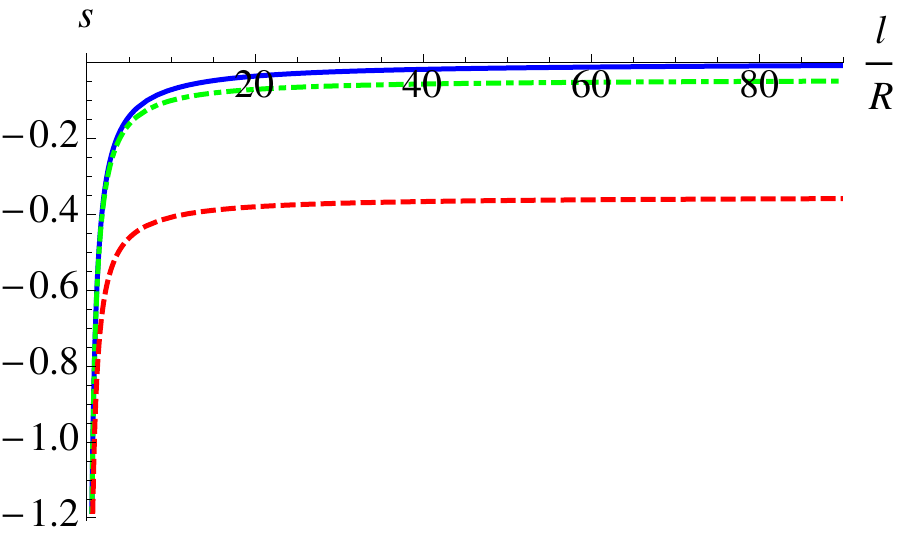}\label{fig:4dstrip}} \hspace{0.5cm}
\subfigure[Disc]{\includegraphics[width=3in]{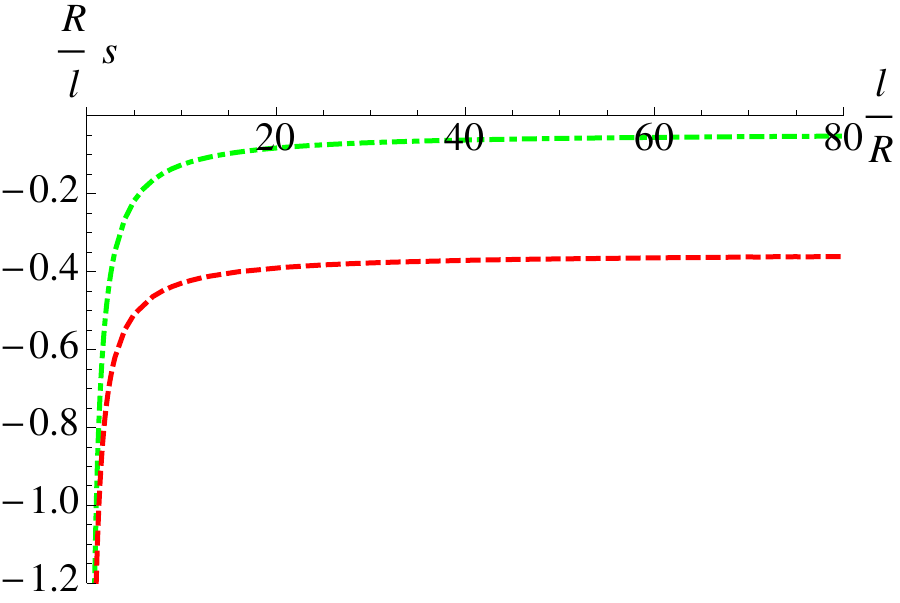}\label{fig:4ddisc}} 
   \caption{\small Results for the strip of width $\ell$ and the disc of radius $\ell$.  Solid blue is the pure AdS$_4$ result, green dot dashed is in a background with $-b_0 = 0.01$, and red dashed is in a background with $-b_0 = 0.001$.}
   \label{fig:4dEE}
   \end{center}
   \end{figure}

\begin{figure}[ht] 
   \centering
   \includegraphics[width=3in]{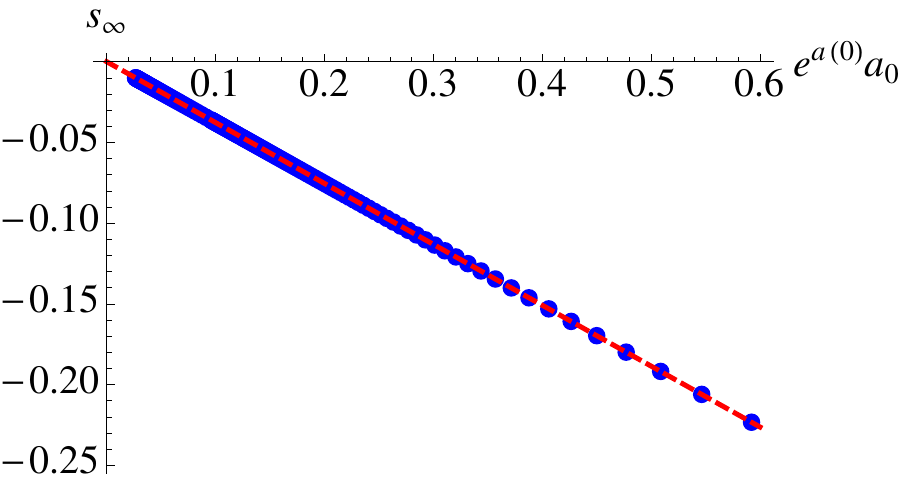} 
   \caption{\small  The coefficient of $\ell$ for large $\ell$.  The blue circles are the numerical results, while the red dashed curve is the linear best fit.}
   \label{fig:4d_drop}
\end{figure}

We follow a similar analysis as before for calculating the entanglement entropy and in light of our detailed calculations presented for AdS$_5$, we will be brief here.
For the strip, we define a quantity $s$ \emph{via}:
\begin{equation}
4 G_N S =\mathrm{Area}= \left( \frac{R^2 L }{\epsilon} + 2 R L s \right) \ ,
\end{equation}
and we plot the result in figure \ref{fig:4dstrip}.  
Similarly for the disc (of radius $\ell$), we define $s$ via:
\begin{equation}
4 G_N S = \mathrm{Area}=2 \pi R^2 \left( \frac{\ell}{\epsilon} + s \right) \ ,
\end{equation}
and we plot the result in figure \ref{fig:4ddisc}.  As expected, there is no new divergent term, and the entanglement entropy has a non--zero coefficient associated with the area of the boundary for large $\ell$.  Furthermore, fitting the large $\ell$ behavior for the disc result to the functional form $a \ell + b$ gives a result for $b$ of approximately  $-0.79$ which is close to the value of $ - R_{\rm IR}^2$, the value we would have predicted in the IR AdS from the thin domain wall analysis (see equation~\reef{final_expand_4_disc} and discussion below it). Using the strip data, we calculate the dependence of the coefficient on the mass and present it in figure \ref{fig:4d_drop}.  We find that the dependence is linear in the mass range explored.  The linear behavior again suggests that we can interpret the coefficient in terms of the mass of the multiplet.  The slope is negative. We expect that we are somewhat away from the mass regime where we connect with the thin  domain wall analysis for physics controlled by the domain wall region.  The regime of larger mass ought to show that the   coefficient increases towards positivity (see equations~\reef{eqt:newarealaw4} and~\reef{eqt:newarealaw4_b}). It is currently too difficult to  extract reliable numerical results in that regime however, and so we cannot test this as we did for the case of one dimension higher.
\section{Conclusions}
We were able to uncover how the holographic entanglement entropy formula encodes a number of field theory features along an RG flow using the key feature of an holographic RG flow, its domain wall. In the thin wall limit where we study an idealized flow between two AdS regions (and hence a flow between fixed point theories) we were able to extract, in various dimensions,  rather pleasing formulae showing how the field theory data appear separated out  according to the various natural scales in the problem.  Using two known RG flows between fixed points, we were able to test our formulae in real examples and found that where we could, several of the features we expected to be robust were confirmed. This included the form of the growth of the entropy upon approach to the IR, which we found defined a new area law supplementing the ones known for the UV behaviour. We were also able to see (in the thin domain wall limit) how the entropy naturally encodes (in its asymptotics)  the   (renormalized) length scale corresponding to the presence of the relevant operator that was switched on.

Our overall goal  is to understand the entanglement entropy  in this holographic setting well enough to use it as another diagnostic tool for  studying the properties of strongly coupled field theories, and so we consider this study of RG flow to be a useful  step along the way. Having analytically characterized, as we have done,  the kinds of behaviour that can appear (in various dimensions),  we expect that holographic  studies  of the entanglement entropy in more complicated theories  will be aided by our results. Many such examples will, by their very nature, be only accessible with numerical approaches, and so analyzing of the results for the entropy may be subtle.  This is where we expect a lot of the intuition here to help bring things sharply into focus, since such features  as the entropy's approach to the end of the flow have now been unpacked in terms of the general features of holographic flow geometries. 

\section*{Acknowledgements}
TA and CVJ would like to thank the Tree for Summer lunchtime inspiration, and Hubert Saluer for conversations. CVJ is supported by the US Department of Energy. CVJ would like to thank the KITP staff for support while some of this manuscript was being prepared, during the workshop ``Holographic Duality and Condensed Matter Physics". TA is supported in part by the US Department of Energy and the USC Dornsife College of Letters, Arts and Sciences. This research was supported in part by the National Science Foundation under Grant No. NSF PHY05-51164.


\providecommand{\href}[2]{#2}\begingroup\raggedright\endgroup

\end{document}